\documentclass[useAMS, usenatbib,fleqn]{mnras}
\usepackage{newtxtext,newtxmath}
\usepackage[T1]{fontenc}
\usepackage{ae,aecompl}
\usepackage[utf8]{inputenc}

\usepackage{amsmath}
\usepackage{amsfonts}
\usepackage{amssymb}

\usepackage{mathtools}	
\usepackage[usenames,dvipsnames]{color}

\usepackage{bm}
\usepackage[british]{babel}

\usepackage{xspace}

\usepackage[normalem]{ulem} %
\usepackage{cancel}

\newcommand{\Gonzalez}{\citetalias{G14}\xspace}
\newcommand{\Lacey}{\citetalias{L16}\xspace}

\newcommand{\ee}{\mathrm{e}}  
\renewcommand{\vec}{\bm} 
\renewcommand{\pi}{\upi}
\renewcommand{\partial}{\upartial}
\newcommand{\rsph}{R}
\newcommand\deriv[2]{\frac{\partial#1}{\partial#2}}
\newcommand\derivsmall[2]{\partial#1/\partial#2}

\newcommand{\vt}{v_0}
\newcommand{\Exp}[1]{{\rm e}^{#1}}
\newcommand{\del}{\partial}

\newcommand{\alp}{\alpha}
\newcommand{\mean}[1]{\overline{#1}}
\newcommand{\meanv}[1]{\overline{\vec{#1}}}

\newcommand{\kin}{_\mathrm{k}} 
\newcommand{\magn}{_\mathrm{m}} 
\newcommand{\ma}{_\mathrm{max}} 
\newcommand{\dd}{\mathrm{d}} 

\newcommand{\Rm}{\mathcal{R}_\mathrm{m}} 

\newcommand{\Rmol}{\kappa}
\newcommand{\Br}{\mean{B}_r}
\newcommand{\Bp}{\mean{B}_\phi}

\newcommand{\muz}{V_Z}

\newcommand{\muG}{\,\mu{\rm G}}

\newcommand{\red}[1]{#1}

\definecolor{webgreen}{rgb}{0,.5,0}
\definecolor{webbrown}{rgb}{.6,0,0}
\definecolor{purple}{rgb}{0.5,0,.5}

\definecolor{orange}{rgb}{1,0.3,0}
\newcommand{\luke}[1]{#1}

\newcommand{\eq}{_\mathrm{eq}}  
  
\newcommand{\Beq}{B\eq}  
                              
\newcommand\vect[1]{\pmb{#1}}                   
\newcommand{\crit}{_\mathrm{c}}  
\newcommand{\turb}{_\mathrm{0}}  
\newcommand{\cs}{c_\text{s}}  
\newcommand{\cm}{\,{\rm cm}}           
\newcommand{\km}{\,{\rm km}}           
             
\newcommand{\pc}{\,{\rm pc}}           
            
\newcommand{\kpc}{\,{\rm kpc}}

\newcommand{\msun}{\text{M}_\odot}                   

\newcommand{\s}{\,{\rm s}}           
         
\newcommand{\Myr}{\,{\rm Myr}}       
\newcommand{\Gyr}{\,{\rm Gyr}}       

\renewcommand{\deg}{^\circ}       

\newcommand{\kms}{\km\s^{-1}}

\newcommand{\erg}{\,{\rm erg}}

\newcommand{\glf}{\textsc{galform}\xspace}

\hypersetup{
    pdfauthor={L.~F.~S.~Rodrigues (orcid.org/0000-0002-3860-0525)
                L.~Chamandy (orcid.org/0000-0003-4935-5550),
                A.~Shukurov (orcid.org/0000-0001-6200-4304),
                A.~R.~Taylor,
                C.~M.~Baugh (orcid.org/0000-0002-9935-9755)},
    pdftitle={Evolution of galactic magnetic fields}
}

\begin{document}
\title[Evolution of galactic magnetic fields]{Evolution of galactic magnetic fields}
\author[L.~F.~S.~Rodrigues et al.]
{\parbox{\textwidth}{
L.~F.~S.~Rodrigues,$^{1}$\thanks{E-mail:
\href{mailto:luiz.rodrigues@ncl.ac.uk}{\texttt{luiz.rodrigues@newcastle.ac.uk} }(LFSR);
\newline \href{mailto:lchamandy@pas.rochester.edu}{\texttt{lchamandy@pas.rochester.edu}} (LC)
}
L.~Chamandy,$^{2,3,4}$
A.~Shukurov,$^{1}$
C.~M.~Baugh$^{5}$ and
A.~R.~Taylor$^{3,4,6}$
}\vspace{0.4cm}\\
\parbox{\textwidth}{
$^{1}$School of Mathematics, Statistics and Physics, University of Newcastle,
Newcastle upon Tyne, NE1 7RU, UK\\
$^{2}$Department of Physics and Astronomy, University of Rochester,  454 Bausch \&
Lomb Hall, Rochester, NY, 14627-0171, USA\\
$^{3}$Astronomy Department, University of Cape Town, Rondebosch 7701, Republic of
South Africa\\
$^{4}$Department of Physics, University of the Western Cape, Belleville 7535,
Republic of South Africa\\
$^{5}$Institute for Computational Cosmology, Department of Physics, University of
Durham, South Road, Durham, DH1 3LE, UK\\
$^{6}$Inter-University Institute for Data Intensive Astronomy, Republic of South Africa
}}

\date{Accepted for publication in MNRAS}
\pagerange{\pageref{firstpage}--\pageref{lastpage}}
\pubyear{2018}
\volume{In press}

\defcitealias{L16}{L16}
\defcitealias{G14}{G14}

\maketitle
\begin{abstract}
  We study the cosmic evolution of the magnetic fields of a large
  sample of spiral galaxies in a cosmologically representative volume
  by employing a semi-analytic galaxy formation model and numerical dynamo solver in tandem.
  We start by deriving time- and radius-dependent galaxy properties using the \glf galaxy formation model,
  which are then fed into the nonlinear mean-field dynamo equations. 
  These are solved to give the large-scale (mean) field
  as a function of time and galactocentric radius for a thin disc, assuming axial symmetry.
  A simple prescription for the evolution of the small-scale (random) magnetic field 
  component is also adopted.
  We find that, while most massive galaxies are predicted to 
  have large-scale magnetic fields at redshift $z=0$, a significant fraction 
  of them is expected to contain negligible large-scale field.
  Our model indicates that, for most of the galaxies containing
  large-scale magnetic fields today, the mean-field dynamo becomes active at $z<3$.
  Moreover, the typical magnetic field strength 
  at any given galactic stellar mass
  is predicted to decline with time up until the present epoch, 
  in agreement with our earlier results. 
  {We compute the radial profiles of pitch angle, and find broad agreement with
  observational data for nearby galaxies.}
\end{abstract}

\label{firstpage}
\section{Introduction}

Galactic dynamo theory has had significant success in explaining the properties of 
the magnetic fields of nearby galaxies
\citep{Ruzmaikin+88, Brandenburg+92, Brandenburg+93, Beck+96, Widrow02, Shukurov05, Shukurov07, Moss+10, Beck2013, Chamandy+16}.
The Square Kilometre Array and other new instruments and techniques
will provide an opportunity to see further back in time at ever larger resolution and sensitivity \citep{Beck+15,Taylor+15,Mao2017}.
Properties of galactic magnetic fields at high redshifts,
albeit in statistically modest samples, have already been probed by 
observations of Faraday rotation in Ly$\alpha$ and  \ion{Mg}{ii} absorption systems 
\citep[][and references therein]{OW95,BMLKD-Z08200,Irwin2013,
FO'SCG14,Farnes2017,Basu2018}
and in a gravitationally lensed late-type galaxy at $z\lesssim1$  \citep{Mao2017}.
\red{The CHANG-ES survey has significantly increased the
sample in the nearby Universe with radio data of 35 nearby edge-on galaxies \citep{Wiegert2015}.}

To provide appropriate theoretical foundations, three crucial ingredients must be
added to dynamo models.
Firstly, models need to take into account that galaxies change over cosmic time, 
and that this can happen suddenly because of galaxy mergers.
Despite the fact that galaxy formation models are well-developed 
\citep{SomervilleDave2015}, dedicated galactic dynamo studies have 
assumed, with rare exceptions, that the underlying galactic parameters are constant 
throughout the evolution of the magnetic field. 
Secondly, galaxies have a wide range of parameters.  
In particular, they vary greatly in size and mass,
with dwarf galaxies having different properties from Milky Way (MW)-like galaxies,
for instance.
Yet most dynamo studies have focused on explaining a generic observed feature, 
or applying a new physical effect, and usually assume the somewhat canonical
parameters that are thought to be roughly suitable for the MW and other
similar-sized spiral galaxies.
Thirdly, theory and observations of galactic magnetic fields need to be advanced beyond the exploration of a relatively small number of specific galaxies to a
\textit{statistical}
analysis of large galaxy samples, both in the nearby Universe and at large redshifts. 
This requires modelling of the statistical distributions of other galactic
properties which affect the dynamo.
Thus, a dynamo model is needed that follows the evolution of galaxies over cosmic
time, is generic enough to be applied to many different types of galaxy,
and can produce a large set of simulated galaxies with a realistic distribution of
properties, magnetic and otherwise.
This is the challenge that this work begins to address.

Previous attempts to model dynamo action in young galaxies focussed on magnetic field
evolution using oversimplified models for the evolution of the host galaxy itself.  
In the first study of this kind,
\citet{BPSS94} employed a nonlinear thin-disc mean-field dynamo model for a generic
spiral galaxy whose disc thickness decreases with time as the galaxy evolves. 
These authors show that the large-scale magnetic field can be amplified to a $\muG$
strength
in 1--$2\Gyr$ and that large-scale magnetic fields in young galaxies are likely to
display global reversals in magnetic field direction that can persist until the
present day.
\citet{Arshakian2009} used order of magnitude estimates based on dynamo theory to
investigate the evolution of magnetic fields in prototypical dwarf, MW-like and
giant galaxies.
Following \citet{BPSS94}, they assumed a mean field dynamo operating on a seed
provided by the fluctuation (or small-scale) dynamo.
Their results indicated that MW-like galaxies could reach $\sim\muG$ fields already
at $z=3$, and that the mean field dynamo typically started operating when the age of
the Universe was of order $1.6\Gyr$.

The first step in incorporating galactic dynamo theory 
into the hierarchical galaxy formation theory was taken by \citet{Paper1}.
There it was assumed that the galactic magnetic field
would  either reach the strength associated with the dynamo steady state in 
galaxies that host a strong dynamo or vanish if the dynamo is sub-critical, 
with galactic properties obtained from a semi-analytic galaxy formation model. 
Because of the assumption of instantaneous adjustment of magnetic field, 
many features of magnetic field evolution remained unresolved. Nevertheless, this
approach allowed one to appreciate how the choice of galaxy formation model
could affect the detailed distribution of magnetic field strengths. 
It was also shown that a fraction of galaxies may not contain active dynamos 
(and thus, large-scale magnetic fields) 
and that the probability of containing a large-scale magnetic field depends on galaxy mass. 
In particular, a large class of satellite galaxies was identified that are unlikely to host
any large-scale magnetic field because of the stripping of their interstellar 
medium (ISM) despite 
possessing kinematic properties favourable for the dynamo action. 
Here we extend this model by resolving the time evolution of magnetic field 
and show that the main conclusions of \citet{Paper1} still hold.

The importance of magnetic fields (and cosmic rays) for galactic evolution is now more
widely appreciated and there have been many recent attempts to include magnetic fields 
in cosmological simulations of forming and evolving galaxies
\citep{Pakmor2013,Pakmor2014,Pakmor2017,Marinacci2017}.
However, these simulations resolve neither the interstellar 
turbulence (which takes place on scales less than about $0.1\kpc$) 
nor the galactic discs (typically, $0.2\text{--}0.5\kpc$).
Our earlier experience with using the velocity field from such a model, 
with adaptive mesh refinement at the highest resolution of $24\pc$, 
as input for a mean-field dynamo model, 
has demonstrated that even this resolution is not sufficient to obtain a robust
model of magnetic field evolution free from extreme sensitivity to parameters
(Agertz et al.\ 2010, unpublished).
Such simulations cannot yet reproduce accurately dynamo action in galactic discs and
haloes.
Analysis similar to that presented here can be performed using
any galaxy formation model but only semi-analytic models provide an opportunity
to consider large galaxy samples.

Here we develop a framework to compute magnetic properties of 
spiral galaxies accounting
for the formation history of individual galaxies produced by a galaxy formation model.
\red{In this paper we consider magnetic fields in galactic discs --  the role of
galactic haloes will be discussed elsewhere.}
For each galaxy, we compute the radial distributions of diffuse gas density
and scale height, as well as the rotation curve. These are used as inputs for
solving the
mean-field dynamo equations, thus modelling the evolution of large-scale magnetic
fields \red{ in the disc}.
We focus on describing the fiducial model and its most general and robust
consequences.
For simplicity, our fiducial model neglects the effects of galactic outflows and
accretion flows on the evolution of magnetic fields.
While we demonstrate that details of magnetic field structure and evolution are
sensitive to specific
properties of the galaxy formation model, we can also identify physical effects
independent of details of a specific galaxy formation model.
A systematic exploration of the parameter space as well as detailed observational
consequences will be discussed in subsequent papers.

The plan of the paper is as follows. Section~\ref{sec:model} presents 
galaxy formation models that we use and our ISM based on them 
(Sections~\ref{sec:angular_velocity}--\ref{sec:density}). Our implementation of the 
fluctuation and mean-field dynamos is discussed in Section~\ref{GMF}.
Section~\ref{results}
contains our results: for magnetic fields in nearby galaxies in 
Section~\ref{sec:local_universe}, for a population of galaxies selected at 
$z=0$ for their significant large-scale magnetic field in Section~\ref{sec:origin}, 
and for a galactic population selected at high redshift in Section~\ref{CEoGM}. 
The radial distributions of large-scale magnetic fields in various 
redshift and galactic mass intervals are discussed in Section~\ref{RMP}. Our results
are put into a broader perspective in Section~\ref{Disc} and summarised in 
Section~\ref{SaC}. Details of calculations can be found in Appendices.

\section{The model}
\label{sec:model}
\subsection{Galactic properties}

We use a semi-analytic model of galaxy formation (SAMGF) to derive the 
evolution of galaxy properties for a large sample of galaxies
of various masses and sizes, consisting of
over a million galaxies in the stellar mass range ${8\lesssim\log(M_\star/\msun)\lesssim12}$. 
In these models, the assembly history of dark matter (DM) structures is first extracted from
a cosmological simulation in the form of a DM halo merger tree. 
The formation and evolution of galaxies is then followed by solving a set of differential 
equations describing the physical processes involved in the star formation and its regulation
\citep[for reviews and references, see][]{Baugh2006,Benson2010,SomervilleDave2015}.
A SAMGF produces a large catalogue of global properties (e.g., disc and bulge stellar masses, 
their gas masses, the disc half-mass radius, assumed to be the same for gas and stars) 
at any given redshift.
From these, using assumptions consistent with those adopted in the SAMGF,
we derive radially-dependent properties of the ISM.

\subsubsection{Galaxy formation models}

We use the \glf SAMGF, introduced by \citet{Cole2000} and continually updated since.
We compare two published versions of \glf: the one described by
\citet[][\Lacey hereafter]{L16} and \citet[\Gonzalez hereafter]{G14}.
Both models use halo merger trees from a Millennium-class $N$-body simulation adopting
{the WMAP7 cosmological parameters \citep{GWAHLB-KTS13}}.
The main difference between the two SAMGFs are the assumptions regarding the stellar
initial mass function (IMF): \Lacey uses the IMF of \citet{Kennicutt} for quiescent
galaxies and assumes a top heavy IMF for starburst galaxies, while \Gonzalez assumes
universally the Kennicutt IMF.

The results presented here correspond to about 7 per cent
of the total volume of the cosmological simulation,
that contains about ${1.4\times10^6}$ tracked galaxies.

\subsubsection{Correction of stellar disc sizes}

\glf generally overestimates the disc sizes of small-mass galaxies and
underestimates the sizes of large-mass galaxies.
We follow the same procedure as
\citet{Paper1} 
and renormalise the half-mass
radii, $r_{1/2}$, of galaxies so that the median of the predicted distribution of
$r_{1/2}$ matches the observed relation between the galactic half-mass radius and
stellar mass at $z=0$ \citep{Lange2016}.
This preserves the dispersion of disc sizes and size evolution
computed by the model, while enforcing realistic final galaxy sizes.

\subsubsection{Morphology of galaxies}

We concentrate on magnetic fields in the discs of spiral galaxies.
We classify a galaxy as spiral in \glf's output if its bulge mass accounts
for less than half of the total stellar mass, $M_\text{bulge}/M_\text{total}<0.5$.

\subsubsection{Angular velocity and rotational shear rate}
\label{sec:angular_velocity}

We reconstruct the rotation curve of each galaxy in agreement with what is used
internally by \glf: we assume that galactic discs are
thin and have an exponential surface density profile, that the galactic bulges 
have a Hernquist {density} profile, and that the dark matter haloes have
an adiabatically contracted NFW profile (see Appendix~\ref{ap:rotation} for details).

While the rotation curve $V(r)$ (where $r$ is the galactocentric distance) obtained 
under these assumptions is realistic away from the galactic centre, the angular velocity 
of rotation $\widetilde\Omega(r)=V(r)/r$ is singular at the rotation axis.
We regularise the angular velocity using
\begin{equation}
 \Omega(r) = \ee^{-(r_\xi/r)^2}\left[\widetilde\Omega(r)-\Omega(r_\xi)\right] + \Omega(r_\xi)\,,
\end{equation}
where we take $r_\xi = 0.1\, r_{1/2}$\;.
The rotational shear rate is obtained as $S(r) = r\del\Omega/\del r$. 
Such a regularisation is compatible with the nature of the mean-field dynamo model 
used to obtain the large-scale magnetic field as it assumes that the gas disc is thin 
and thus only applies at $r\gtrsim r_\xi$.

\subsubsection{ISM turbulence}
\label{sec:ism}
The turbulence in the ISM is mostly driven by supernova (SN) explosions.
We assume that the root-mean-square turbulent speed $v\turb$ is equal to the
sound speed $\cs$ in the warm gas ($T\simeq 10^4\,\text{K}$),
\begin{equation}
  v\turb = \cs = 10 \kms\,.
\end{equation}
As discussed in Section 2.2.1 of \citet[][]{Paper1}, these quantities
\red{are expected to}
vary little between and within spiral galaxies.
\red{This assumption agrees with recent observations in \ion{H}{i} and CO
\citep[see][]{Mogotsi2016}.}

We assume that the turbulent scale is controlled by the size of a supernova remnant 
when its expansion velocity reduces to the sound speed in the warm gas, which is 
of order $0.1\kpc$.
However, the turbulent scale is limited from above by the half-thickness of the galactic
gaseous disc because larger supernova remnants and their clusters break through the disc
into the halo. Thus, we adopt for the turbulent scale
\begin{equation}
 l_0(r) = \min\left[ 100\pc, h(r)\right]\,,
\end{equation}
where $h(r)$ is the scale height of the diffuse gas disc.

\subsubsection{Diffuse gas density and scale height}
\label{sec:density}

\glf does not include a detailed description of the ISM. 
Thus, the radial profiles of the gas volume density $\rho(r)$ and
scale height $h(r)$ have to be estimated from the simulated galaxy properties
such as the total stellar mass $M_\star$ and the total gas mass of the galactic 
disc $M_\text{g}$.

As is done internally in \glf for the stellar disc, the surface density of 
the gaseous disc is assumed to have an exponential radial profile with the same scale length $r_\mathrm{s}$, 
so for both stars and gas the mass surface densities are adopted as
\begin{equation}
 \Sigma_\text{g}(r) = \frac{M_\text{g}}{2\pi r_\mathrm{s}^2}
\ee^{-r/r_\mathrm{s}}\quad\text{and}\quad
 \Sigma_\star(r) = \frac{M_\star}{2\pi r_\mathrm{s}^2}
\ee^{-r/r_\mathrm{s}}\,,\label{eq:Sigma}
\end{equation}
 where $r_\mathrm{s}$ is the scale length (which is the same for stars and total gas).

At the level of spatial resolution available in any galaxy formation model, 
the multi-phase structure of the ISM can only be allowed for in an approximate manner.
The large-scale dynamo appears to operate in the warm phase 
\citep{Evirgen2017} because the cold gas occupies a negligible fraction of the 
disc volume whereas the hot phase is unsteady and leaves the disc for the gaseous halo
in a fountain outflow or a wind on time scales shorter than the time scale of the mean-field
dynamo. Therefore, we adopt for the ISM the density and temperature characteristic of the 
warm interstellar gas. 

Most of the mass of the cold interstellar gas is in molecular form whereas the hot gas
contributes negligibly to the gas mass. The total gas mass is therefore separated into
the diffuse (warm) and molecular (cold) phases
following a procedure similar to that used by \citet{Lagos2011gas}.
We compute the ratio of molecular to diffuse gas surface densities
$\Rmol=\Sigma_\text{m}/\Sigma_\text{d}$ using the empirical relation
found by \citet{BlitzRosolowsky2004,BlitzRosolowsky2006} which relates $\Rmol$
to the total gas surface density of $\Sigma_\text{g}$ and that of stars, 
$\Sigma_\star$ (see Appendix~\ref{ap:fmol}).
From $\Rmol$, the surface densities of diffuse and molecular gas components
are obtained as, respectively,
\begin{equation}
    \Sigma_\text{d}(r) = \frac{\Sigma_\mathrm{g}(r)}{1+\Rmol(r)}
    \quad\text{and}\quad
    \Sigma_\text{m}(r) = \Rmol(r)\Sigma_\mathrm{d}(r)\,.
\end{equation}
Since $\Rmol$ is a function of galactocentric distance $r$, the diffuse and molecular gas 
phases have distinct radial scale lengths that differ from that of the total gas density.

We adopt exponential density profiles in the coordinate $Z$ perpendicular to the disc surface
for the diffuse and molecular gas as well as the stars to obtain the volume 
mass densities:
\begin{equation}
 \rho_\text{d} = \rho_{\text{d}0}\ee^{-Z/h_\text{d}}\,,
\quad
 \rho_\text{m} = \rho_{\text{m}0}\ee^{-Z/h_\text{m}}\,,
\quad
 \rho_\star = \rho_{\star0}\ee^{-Z/h_\star}\,,\label{eq:vert_exp}
\end{equation}
where $\rho_{\text{d}0}$, $\rho_{\text{m}0}$ and $\rho_{\star0}$ are the corresponding 
mid-plane values.
Assuming that the ISM is in the state of statistical hydrostatic equilibrium, this allows us
to derive the mid-plane total gas pressure profile as (see Appendix~\ref{ap:P} 
for the derivation)
\begin{align}\label{eq:P}
 P(r) =& \frac{\pi}{2} G \,\Sigma_\text{d}(r) \left\{ \Sigma_\text{d}(r)
                \left[1 + \frac{2\,h_\text{d}(r)R_\text{m}(r)}{h_\text{m}+h_\text{d}(r)}
                \right]
                +  \Sigma_\star(r)
                \frac{2\,h_\text{d}(r)}{h_\star+h_\text{d}(r)} \right\}\nonumber\\
       &\quad + h_\text{d}(r)\Sigma_\text{d}(r)\left[\tfrac{3}{2}
        \Omega^2_\text{b}(r)+\Omega_\text{b}(r)\;S_\text{b}(r)\right]\nonumber\\
       &\quad + h_\text{d}(r)\Sigma_\text{d}(r) \Omega_\text{dm} 
	\left[\tfrac32         \Omega_\text{dm}(r)+S_\text{dm}(r)\right],
\end{align}
where $\Omega_\mathrm{i}(r)$ and $S_\mathrm{i}(r)=r\del\Omega_\mathrm{i}/\del r$
are the angular speed and shear rate of the component $\text{i}=\text{d,m,$\star$,b,dm}$,
for the diffuse and molecular gas, stellar disc, galactic bulge and dark matter, respectively 
(see Appendix~\ref{ap:rotation} for details).

The stellar scale height is observed to be approximately constant with $r$ and
related to the radial scale length as $h_\star= \beta_\star r_\text{s}$ with
$\beta_\star\approx0.1$ \citep{Kregel2002}.
The scale height of the molecular gas in the Milky Way is also approximately
constant with $h_\text{m,MW}\approx 80\pc$ \citep{Cox2005,HeyerDame2015}.
We assume that this applies to other galaxies, with their molecular
gas scale heights scaling with their disc sizes (analogously to the stellar disc):
$h_\text{m}=\beta_\text{m} r_\text{s}$, with $\beta_\text{m}\approx0.032$
and $r_{s}\approx 2.5\kpc$ in the MW \citep{Licquia2016}.

Altogether, by accounting for all sources of gravitational support, we have expressed 
in equation~\eqref{eq:P} the mid-plane total gas pressure in terms of the
scale height, molecular fraction, surface densities and rotation curves. In order to 
derive the scale height of the diffuse gas, we represent the total pressure
(the sum of thermal, turbulent, magnetic and cosmic-ray contributions), as a multiple 
of the turbulent pressure,
\begin{equation}
 P(r) = \zeta \rho_\text{d}(r) v\turb^2\,,
 \label{eq:Pleft}
\end{equation}
where $\zeta \approx {1.1}$ 
is obtained by accounting for the
non-thermal pressure components assuming various types of equipartition between 
them (see Appendix~\ref{ap:Pgas} for the derivation).
Equations~\eqref{eq:P} and \eqref{eq:Pleft} 
express the total mid-plane gas
pressure in terms of the surface and volume gas densities, respectively, thus allowing us
to obtain the scale height of the diffuse gas,
\begin{equation}
  h_\text{d}(r) = \frac{\Sigma_\text{d}(r)}{2\rho_\text{d}(r)}\label{eq:hSigma}\,.
\end{equation}
From this point on, unless otherwise stated, we will omit the subscripts `d' in 
the density and scale height of the diffuse gas.
As discussed below, this leads to a form of a galactic flared disc that is fully
compatible with \ion{H}{i} observations in the MW.

\subsection{Galactic magnetic fields}\label{GMF}

Turbulent dynamo theory can be conveniently separated into fluctuation dynamo theory,
which describes the generation of random magnetic fields by random flows on scales smaller
than the outer scale of turbulence (the random or small-scale field),
and mean-field dynamo theory, which explains the origin of magnetic fields
at scales larger than those of turbulence (the mean, large-scale or regular 
magnetic field).
The mean-field dynamo action also leads to the generation of
small-scale magnetic fields but these are uniquely related to the mean magnetic field.
The two dynamo mechanisms can act independently and their interaction remains a matter
of intense study \citep[][and references therein]{Brandenburg+Subramanian05a}.
Both dynamos are threshold phenomena: the fluctuation dynamo can maintain a random
magnetic field when the magnetic Reynolds number $\Rm$ exceeds a certain critical value
of order 100 depending on the detailed nature of the plasma flow, whereas the mean-field 
dynamo amplifies and then sustains a large-scale magnetic field when, in the simplest case, 
the dynamo number (depending on the rotation and its shear rates, as defined below) exceeds 
a critical value of order 10 in a thin disc.
For galaxies, the magnetic Reynolds number $\Rm$ exceeds the critical value by 
several orders of magnitude, and thus interstellar turbulence can amplify an arbitrarily weak
random magnetic field exponentially on a time scale comparable to or shorter than 
the eddy-turnover time $l\turb/v\turb$ of energy-carrying eddies, 
of order $10\Myr$ in the Solar neighbourhood.
On the other hand, the dynamo number in spiral galaxies is not far from its threshold value,
and the large-scale dynamo may or may not be active in a given galaxy depending
on its rotation properties and the thickness of its gas layer. For both dynamo mechanisms,
the magnetic energy density in a steady-state is of the order of the turbulent energy 
density. We provide below a quantitative discussion of these mechanisms.

In what follows, we represent magnetic and velocity fields as the sums of large-scale
and random parts, denoted with a bar and lower-case symbols, respectively:
\begin{equation}\label{BbVv}
\vec{B}=\meanv{B}+\vec{b}\,,\qquad \vec{V}=\meanv{V}+\vec{v}\,,
\end{equation}
where $b\turb$ and $\vt$ are the respective root-mean-square values of the random parts.

\subsubsection{Small-scale magnetic field}
\label{sec:ss}
{Random magnetic fields in the ISM of spiral galaxies are produced by the fluctuation 
dynamo action and, independently, together with the mean magnetic field as a part of the 
mean-field dynamo mechanism \citep[e.g., Section 7.4.2 in][]{Shukurov07}.}
As in \citet{Paper1}, we assume that the energy density of the random magnetic
field is a fraction $f_B$ of the interstellar turbulence energy density,
\begin{equation}\label{Beq}
  b\turb=f_BB\eq\,,   \quad  B\eq=(4\pi\rho)^{1/2} \vt\,,
\end{equation}
where $\rho$ is the diffuse gas density.
Our fiducial model takes $f_B=0.5$
\citep{Brandenburg+Subramanian05a,Basu+Roy13,Paper1,Kim+Ostriker15}.
Theory and simulations of nonlinear fluctuation and mean-field dynamos still cannot
provide a more detailed estimate than this.

\subsubsection{Mean-field dynamo equations}
We use the standard equations of the nonlinear turbulent mean-field dynamo with 
simplifications appropriate to the thin discs of spiral galaxies. In particular, 
differential rotation is assumed to be the dominant source of the azimuthal magnetic 
field \citep[the $\alpha\omega$ approximation --][]{Ruzmaikin+88} and derivatives of 
$\meanv{B}$ along the $Z$-axis perpendicular to the disc mid-plane are replaced by 
appropriate ratios involving the scale height of the diffuse gas
\citep[the no-$Z$ approximation --][]{Subramanian+Mestel93,Moss95,Phillips01,Chamandy+14b}. 
Apart from differential rotation, we include the outflow at a speed $V_Z$. 
In cylindrical coordinates $(r,\phi,Z)$ centred at the galactic centre with the $Z$-axis
aligned with the galactic rotation axis, equations for the radial and azimuthal components
of the mean magnetic field reduce to
\begin{align}
  \label{dBrdt}
    \deriv{\Br}{t}=&-\frac{\muz \Br}{2h}
                    -\frac{2\alp\Bp}{\pi h}
                    +\eta\left\{-\frac{\pi^2\Br}{4 h^2}
                    +\deriv{}{r}\left[\frac{1}{r}\deriv{(r\Br)}{r}\right]\right\}\,,\\
  \label{dBpdt}
    \deriv{\Bp}{t}=&-\frac{\muz\Bp}{2h} +S\Br
         +\eta\left\{-\frac{\pi^2\Bp}{4h^2}
         +\deriv{}{r}\left[\frac{1}{r}\deriv{(r\Bp)}{r}\right]\right\}\nonumber\\
        &+\deriv{\eta}{r}\left(\deriv{\Bp}{r}+\frac{\Bp}{r}\right)+f\,,
\end{align}
where \red{$h(r)$ is the scale height profile obtained in Section~\ref{sec:density},
$S(r)$ is the shear rate profile, and }
$\alp\red{(r)} = \alp\kin\red{(r)}+\alp\magn\red{(r)}$, with $\alp\kin$ representing the effect of the
background mean helicity of interstellar turbulence and $\alp\magn$ the modification 
of the mean helicity induced by the magnetic field 
\luke{\citep{Pouquet+76,Kleeorin+Ruzmaikin82,Gruzinov+Diamond94,Bhattacharjee+Yuan95}}. 
The following equation for $\alp\magn$ 
closes the system of equations by allowing for the nonlinear back-reaction 
of the mean magnetic field on the turbulence \luke{\citep[e.g.][]{Kleeorin+02,Blackman+Field02,Shukurov+06}}:
\begin{align}
  \label{dalpha_mdt}
    \deriv{\alp\magn}{t}=&-\frac{2\eta}{l\turb^2B\eq^2}\left[\alp(\Br^2+\Bp^2)
                     +\frac{3\eta}{h}\sqrt{\frac{|D|}{\pi^{3}}}\Br\Bp \right]
                     -\frac{\muz\alp\magn}{h}\nonumber\\
                     &+R_\kappa\eta\left[-\frac{\pi^2\alp\magn}{h^2}
               +\frac{1}{r}\deriv{}{r}\left(r\deriv{\alpha\magn}{r}\right)\right]
                                +R_\kappa\deriv{\eta}{r}\deriv{\alp\magn}{r}\,.
\end{align}
The $Z$-component of $\meanv{B}$ can be obtained either from the corresponding
component of the dynamo equation (not shown) or, equivalently, from the solenoidality 
condition $\nabla\cdot\meanv{B}=0$. Derivation and detailed discussion of these equations 
and their solutions can be found in \citet{Chamandy+13a,Chamandy+14b} and \citet{Chamandy16}. 

The source term $f$ in equation~\eqref{dBpdt} (its form is derived in 
Section~\ref{sec:floor}) is designed to model the statistical contribution of the 
random magnetic field to the mean-field equation: the mean value of $\vect{b}$ only
vanishes in infinite space but remains significant in the finite volume of a galactic disc.

The background $\alpha$-effect $\alpha\kin$ is estimated as
\begin{equation}
  \alpha\kin = \min\left(\frac{l\turb^2(r) \Omega(r)}{h(r)}, v\turb\right)\,,\label{eq:alpha_kin}
\end{equation}
where the upper limit $v\turb$ can be reached in the central parts of galaxies where
$\Omega$ is large and $h$ is small, but this rarely happens.

The turbulent magnetic diffusivity is estimated from the mixing length theory as
\begin{equation}
  \label{eta}
  \eta = \tfrac{1}{3} l\turb v\turb\,.
\end{equation}
The microscopic diffusion has been neglected in equations~\eqref{dBrdt}--\eqref{dalpha_mdt}, 
which is appropriate given that $\Rm$ is large.
The diffusive flux of magnetic helicity {within the disc and} through the disc surface
into the halo is characterised by the parameter $R_\kappa$
which we assume to be equal to unity corresponding to equal turbulent diffusivities
of magnetic field and helicity \citep[for details, see][]{Chamandy+14b}.
\red{Since the diffusive flux of magnetic helicity is expected to dominate over the
advective flux in the disc \citep[see Section~4.7 of][]{Paper1}, we neglected the effect of outflows
on the magnetic field in the fiducial model we present here.
}

The intensity of the mean-field dynamo action in the $\alpha\omega$ approximation
is controlled by the dynamo number $D$, defined in terms of the nonlinear $\alpha$-effect
$\alp=\alp\kin+\alp\magn$ as 
\begin{equation}\label{D}
  D = \left(\alp\kin+\alp\magn \right) \frac{h^3 S}{\eta^2}\,.
\end{equation}
We also use the `kinematic' dynamo number defined similarly but for the $\alpha$-effect
unaffected by magnetic field,
\begin{equation}\label{Dkin}
  D\kin=\alp\kin\frac{h^3S}{\eta^2}\,.
\end{equation}
{$D\kin$ characterises the ability of a galaxy to launch the mean-field dynamo}
{and contributes to the control of the steady-state strength of the magnetic field},
 whereas $D$ is approximately equal to $D\kin$ when $B\ll B\eq$,
and approaches the critical (threshold) values as the dynamo saturates.
Both $D$ and $D\kin$ are functions of $r$. 
For the dynamo to be active in a given range of $r$,  $D(r)/D\crit(r)$ should exceed unity. 
Since $\alp\kin$ and $\alp\magn$ normally have opposite signs, both
$|\alp|=|\alp\kin+\alp\magn|$ and $|D|$ decrease as magnetic field grows, leading to
saturation of the dynamo action and the establishment of a quasi-steady state in which 
$D(r)/D\crit(r)\approx1$. In this state, the magnetic field is subject to secular variation 
due to the  evolutionary variation of galactic parameters. When $|D|$ decreases below the 
threshold value $|D\crit|$ of order 10 in a thin disc, magnetic field growth is halted and 
further  decrease in $|D|$ leads to the decay of the mean magnetic field at a rate
comparable to but lower than the inverse turbulent diffusion time across the disc. 
The term $f$ in equation~\eqref{dBpdt} prevents the mean magnetic field from decreasing below 
the (low) strength that corresponds to the average strength of the random magnetic field in
the finite galactic volume as discussed in Section~\ref{sec:floor}. This weak mean magnetic 
field serves as a seed for the dynamo action should $|D\kin|$ become supercritical again due 
to the evolution of the galaxy.

Our dynamo model allows for periods of active dynamo action as well as periods of 
magnetic field decay depending on the nature of the galactic evolution. In particular,
we assume that major mergers (i.e., those where the ratio of the merging stellar masses 
exceeds 0.3)
lead to a dispersal of the galactic
gaseous discs and their magnetic fields. The mean-field dynamo action can resume after 
such a merger if $|D\kin|>|D\crit|$ in the newly formed galaxy but it starts with a weak 
seed magnetic field.

\subsubsection{Random magnetic field and the mean-field dynamo}
\label{sec:floor}

The spectrum of the fluctuating magnetic field discussed in Section~\ref{sec:ss}
has a tail extending to small wave numbers. Moreover, averaging of the
magnetic fluctuations in a finite volume of the galactic disc does not vanish but rather
scales as $N^{-1/2}$, where $N$ is the number of correlation cells of $\vect{b}$ in the
disc volume \citep[Section VII.14 in][]{Ruzmaikin+88}. This statistical
large-scale tail acts as a seed for the mean-field dynamo action that is supplied constantly
and throughout the galactic disc as long as interstellar turbulence keeps generating 
random magnetic fields. Therefore, it allows an evolving galaxy to launch the mean-field 
dynamo action again after a possible period of inactivity.
For simplicity, we only include the corresponding source term $f$ into the azimuthal 
component of the dynamo equation \eqref{dBpdt}. The magnitude of this seed has been
estimated by \citet{Ruzmaikin+88} as
\begin{equation}
  \label{B0amp}
   \widetilde{B}\simeq\tfrac{1}{3}\frac{b\turb}{N^{1/2}}\frac{l}{\Delta r}\,,
\end{equation}
where $\Delta r$ ($\simeq2\kpc$) is the radial width of the leading eigenfunction 
of the mean-field dynamo equation (close to the galactocentric radius where the
rotational shear is maximum). This estimate allows for the fact that $\nabla\cdot\vect{b}=0$.
The number of spherical turbulent correlation cells of a radius $l\turb$ in a cylindrical 
shell of a radius $r$ ($\gg\Delta r$), width $\Delta r$ and height $2h$ is estimated as
\begin{equation}
  N\simeq\frac{4\pi hr\Delta r}{(4/3)\pi l\turb^3}=\frac{3hr\Delta r}{l\turb^3}\,.
\end{equation}
To avoid the formal singularity at $r\to0$, we introduce an exponential truncation
at small radii (that only becomes important at $r\ll\Delta r$) and adopt
\begin{equation}
   \widetilde{B}(r) = \tfrac{1}{3}\frac{b\turb(r)}{N^{1/2}(r)}
                    \frac{l\turb(r)}{\Delta r}\Exp{-\Delta r/(2r)}\,.
                    \label{eq:B0}
\end{equation} 

\begin{figure*}
  \centering
  \includegraphics[width=0.97\columnwidth]{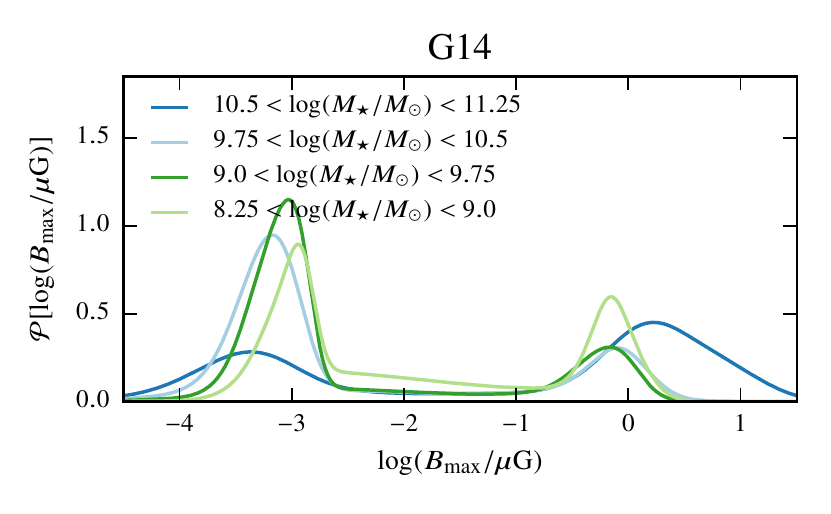}\;\;
  \includegraphics[width=0.97\columnwidth]{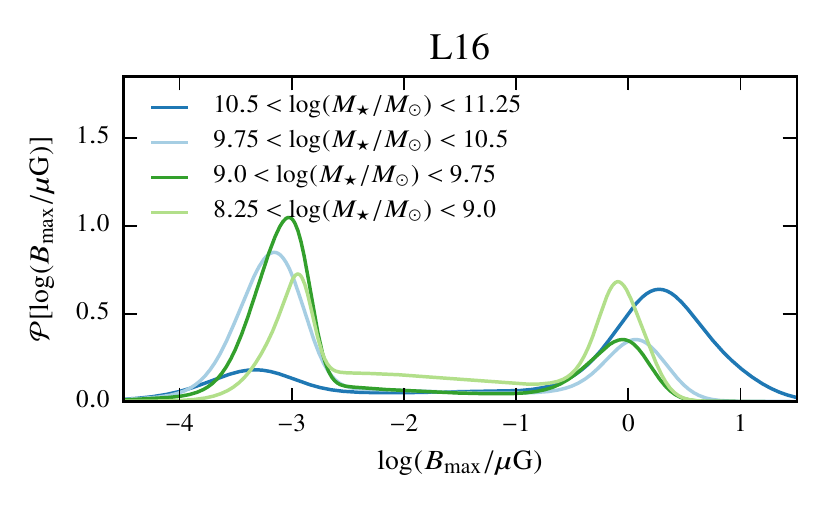} \\ \smallskip
  \includegraphics[width=0.97\columnwidth]{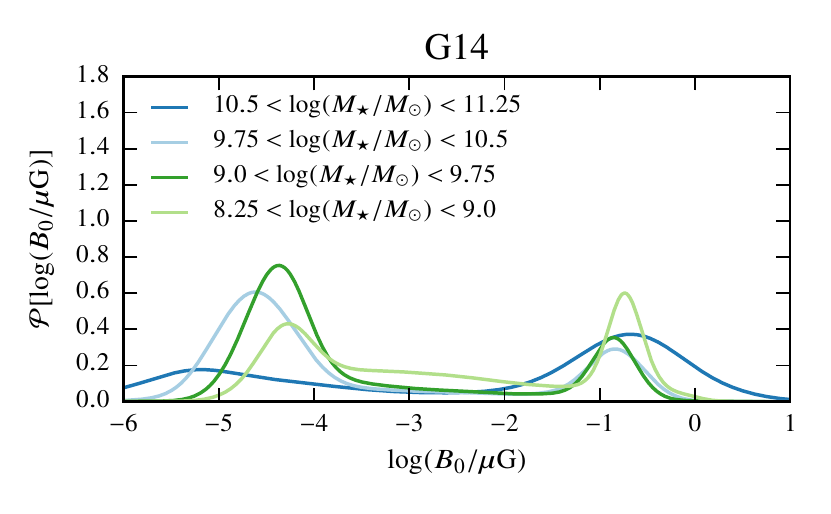}\;\;
  \includegraphics[width=0.97\columnwidth]{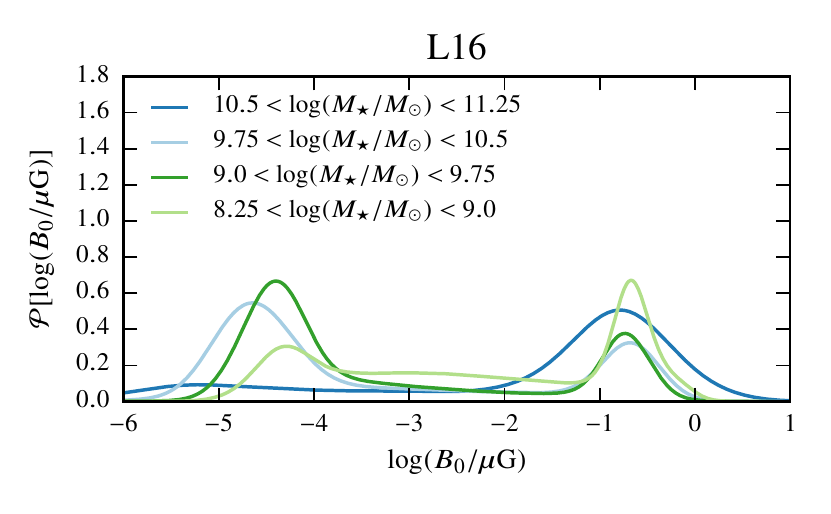}
  \caption{Probability density functions of $\log({B}\ma)$ and $\log({B}_0)$
           at $z=0$ for spiral galaxies in various mass ranges.
           The left- and right-hand columns show the results for the \Gonzalez and
           \Lacey models.
           }
  \label{fig:PDFs}
\end{figure*}

\begin{figure}
  \centering
  \includegraphics[width=0.95\columnwidth]{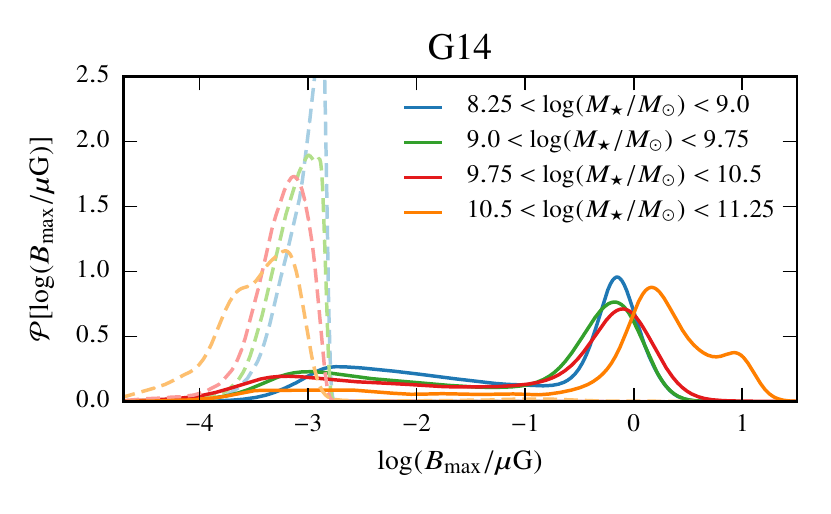}\\ \smallskip
  \includegraphics[width=0.95\columnwidth]{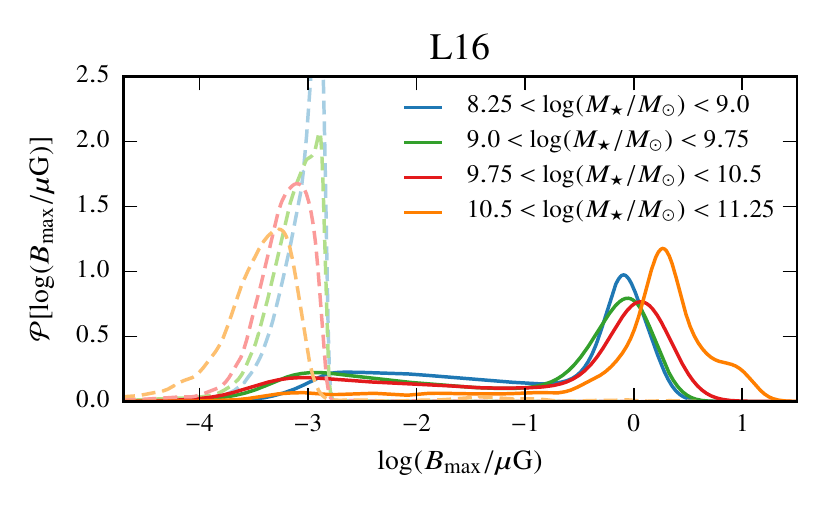}
  \caption{Probability density functions of $\log({B}\ma)$ at $z=0$ in various
           galactic stellar mass ranges for the \Gonzalez (top panel) and
           \Lacey (bottom panel) galaxy formation models.
           Dashed curves show spiral galaxies with subcritical dynamos at
           the radius of maximum field strength, i.e., 
           $D\kin(r\ma)/D\crit(r\ma)<1$, while
           solid curves correspond to galaxies with active dynamos,
           $D\kin(r\ma)/D\crit(r\ma)>1$.}
  \label{fig:PDFsActPas}
\end{figure}

For $f$, we adopted, after some experimentation, 
a form that {becomes important only for a strongly subcritical dynamo.
We choose this term such that, if the dynamo is sub-critical, $\Bp$ decays towards
a steady state with $\Bp\approx\widetilde{B}$. To obtain such a form for $f$, we
simultaneously solve equations~\eqref{dBrdt} and \eqref{dBpdt} analytically assuming 
$\del/\del t=0$, $\alpha=\alpha\kin$ and $\del/\del r=0$ (as radial diffusion is 
sub-dominant in a thin disc). Then $\Bp=\widetilde{B}$ leads to
\begin{equation}
  \label{f}
  f= \frac{\pi^2\eta}{4h^2}\left(1+\frac{R_U}{\pi^2}\right)\left(1-\frac{D}{D\crit}\right) \widetilde{B}\,,
\end{equation}
where
\begin{equation}
  R_U= \frac{\muz h}{\eta}\,,
  \quad
  D\crit= -\left(\frac{\pi}{2}\right)^5\left(1+\frac{R_U}{\pi^2}\right)^2
  \label{eq:Dc}
\end{equation}
are, respectively, the dimensionless outflow velocity (or the corresponding 
turbulent magnetic Reynolds number), 
and the critical dynamo number. 
We have verified that this prescription causes $\Bp(r)$ to converge to a profile
very close to that of $\widetilde{B}(r)$. If the dynamo is only slightly subcritical, 
then $\Bp$ can be a few times smaller in magnitude than $\widetilde{B}$ in the steady 
state, due to the radial diffusion neglected in equation~\eqref{f}.

It is unclear how to rigorously prescribe the sign of $\widetilde{B}$: it can
vary in both time and space. This variation has observational significance since it can
give rise to magnetic field reversals. We applied the following prescription for the choice
of the sign of $\widetilde{B}$: it is chosen randomly 
along $r$ remaining sign-constant within an annulus $10 l\turb$ wide, and changes
randomly over every time interval $10\tau\turb=10 l\turb/v\turb$. For the parameter values
typical of the Solar neighbourhood, $b\turb\simeq5\muG$, $l\turb\simeq0.1\kpc$ and 
$l\turb/v\turb\simeq10\Myr$, $N\simeq3\times10^4$ and 
$\widetilde{B}\simeq5\times10^{-4}\muG$.

\subsubsection{Numerical implementation}

From the output of the galaxy formation model (\glf), 45 fixed-redshift snapshots are 
extracted, corresponding to the properties of $1.4\times10^6$
galaxies from $z=6$ to $z=0$ (evenly spaced in the logarithm of the 
cosmological scale factor, which
corresponds to time intervals between $100$ and $300\Myr$).
The galaxy properties are taken as approximately constant between two consecutive 
snapshots, 
and for each interval the angular velocity, density and scale height of diffuse gas are 
computed for each galaxy as functions of the galactocentric distance $r$.

All radially dependent quantities are computed on an evenly spaced grid spanning the 
range $0< r \leq 2.7 r_{1/2}$, where $r_{1/2}$ is the half-mass radius of the
galaxy {(including the baryon mass alone)}. If it is the first snapshot 
or if $r_{1/2}$ did not increase since the previous snapshot, we use the number of 
the grid points $n_r=55$.
If the galaxy 
had increased in size, the number of grid points is temporarily increased so
that $r=2.7 r_{1/2}$ can be accommodated using the same radial resolution as in the
previous snapshot. After finishing the calculation for a given snapshot, the augmented grid
is interpolated back into an $n_r$-point grid.
For a typical value $r_{1/2}=\red{1}0\kpc$, the standard grid separation,
about $\delta r=0.\red{2}\kpc$, is comfortably smaller than the radial scales of the angular
velocity, disc scale height, large-scale magnetic field and the seed field $\widetilde{B}$. 
Since our sample contains a large number of galaxies, significantly larger values of $n_r$ 
can be computationally unaffordable.

Over each time interval between the snapshots, equations~\eqref{dBrdt}--\eqref{dalpha_mdt} 
are solved numerically using a third order Runge--Kutta time stepping and sixth order 
spatial derivatives \citep{Brandenburg2003}.
The time step $\delta t$ is chosen dynamically to satisfy the following 
Courant--Friedrichs--Lewy condition based on the advection time scale and magnetic 
diffusion time across the minimum disc scale height:
\begin{equation}
       \delta t = 0.16\min\left[\frac{\delta r}{V}\,,\ 
                        \frac{\underset{r}{\min} h^2(r)}{\eta} \right].
\end{equation}
We assume $\mean{B}_\phi = \mean{B}_r = 0$ at both the inner and outer boundaries.
We have tested alternative outer boundary conditions
$\derivsmall{B_\phi}{r}=\derivsmall{B_r}{r}=0$ and found negligible impact on the results
in the test runs.

\section{Results}\label{results}

While our model produces full radial profiles of magnetic properties for each
galaxy, it is convenient to have a few diagnostic quantities which can characterise
the magnetic field in each galaxy. For this purpose we use the 
maximum strength of the large-scale magnetic field
and the corresponding galactocentric distance $r\ma$,
\begin{equation}
  {B}\ma=\max_r |\vec{\mean{B}}(r)| =|\vec{\mean{B}}( r\ma )|\,, \label{eq:defBmax}
\end{equation}
as well as the root-mean-square (rms) large-scale field strength $B_0$,
\begin{equation}
{B}^2_0 = \frac{\int_0^{\infty}|\vec{\mean{B}}(r)|^2 h(r)\, r\;\dd r}{
              \int_0^{\infty} h(r) \,r\;\dd r}\,. \label{eq:defBeavg}
\end{equation}

\subsection{Magnetic fields strength in nearby galaxies}
\label{sec:local_universe}

Before discussing magnetic fields in high-redshift galaxies, we evaluate our model
at $z=0$ where the results can be compared with observations of nearby galaxies. 
Figure~\ref{fig:PDFs} shows the probability distribution functions (PDFs) of magnetic field 
strengths in four galactic stellar mass ranges. For both galaxy formation models used 
and for every mass range, there are
two populations, one with mode ${B}\ma \simeq 1\muG$ and another with mode
${B}\ma \simeq 10^{-3}\muG$. The large-scale field strength of the latter population is
close to the seed field strength $\widetilde{B}$ described in Section~\ref{sec:floor}. 
The galaxies in the other population display large-scale magnetic 
field strengths that are in broad agreement with the observations.
The rms magnetic field, \red{$B_0$}, similarly has a bimodal probability distribution.
The two populations are clearly separated by a wide minimum at 
$10^{-3.5}\lesssim {B}_0\lesssim 10^{-\red{1}.5}\muG$ and we select, for convenience,
${B}_0=0.05\muG$ as the field strength separating them.
The fraction of galaxies with
$0.05\lesssim{B}_0\lesssim\red{0.1}\muG$ is negligible, so, for our present purposes,
galaxies with ${B}_0>0.05\muG$ can be considered to host significant large-scale
magnetic fields.

In Fig.~\ref{fig:PDFsActPas}, the PDFs of ${B}\ma$ are shown separately
for galaxies with active dynamos (i.e., $D\kin/D\crit>1$ at ${r=r\ma}$) and
galaxies incapable of sustaining dynamo action (${D\kin/D\crit<1}$ at ${r=r\ma}$),
confirming that the population with the lower field strength in Fig.~\ref{fig:PDFs} 
hosts sub-critical mean-field dynamos.

\begin{figure}
  \centering
  \includegraphics[width=1.05\columnwidth]{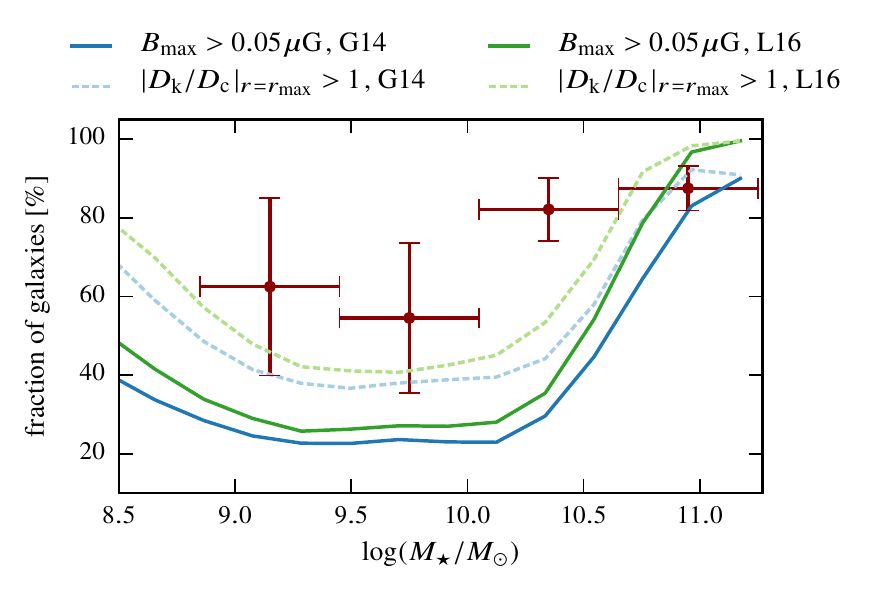}
  \caption{Fraction of spiral galaxies containing large-scale magnetic field and/or active
           mean-field dynamos at $z=0$, showed for both \Lacey (green) and 
           \Gonzalez (blue) models.
           The likelihood that a galaxy hosts a large-scale magnetic fields depends
           on its stellar mass, with less than 40 per cent of the galaxies with stellar masses
           about $4\times10^9\msun$ containing significant magnetic fields, with the line styles
           explained in the legend above the figure frame.
           The red circles with error bars show the fraction of observed galaxies
           containing large-scale magnetic fields in the compilation of 
           \citet{Beck2013} and stellar mass data from the
           S4G catalogue and other sources (see text for details). The vertical bars
           represent the Poisson errors and the horizontal bars show the width of the 
           logarithmic mass range used.
           Note that mass ranges are different from those used for the model 
           results in other figures.
          }
  \label{fig:fractions}
\end{figure}

There appears to be a connection between the galactic stellar mass and magnetic field strength: 
most galaxies in the largest-mass range contain strong magnetic fields at $z=0$ while 
galaxies of a smaller mass host negligible fields; 
however, this dependence is not monotonic and even
smaller galaxies often contain stronger
magnetic fields. This is demonstrated in Fig.~\ref{fig:fractions}, where we show
the fraction of galaxies with a significant field strength exceeding $\widetilde{B}$, 
${B}\ma>0.05\muG$, as a function of the stellar mass at $z=0$.
For the \Lacey model (solid green) this fraction decreases below 40 per cent
for $M_*\simeq (1\text{--}20)\times10^9\msun$ 
and increases for larger and smaller masses.
For the \Gonzalez model (solid blue{,  in Fig.~\ref{fig:fractions}}), the trend is similar, but the
fraction of galaxies with a significant large-scale magnetic field is somewhat smaller.
In the same figure, we show the fraction of galaxies
where the dynamo is active ($D\kin/D\crit>1$) at $r=r\ma$
depends on the stellar mass. The behaviour is similar to that described above but the 
fractions are larger, indicating that some of the galaxies did not have enough time to 
reach the $0.05\muG$ threshold even though they currently host an active dynamo.

The circles with error bars in Fig.~\ref{fig:fractions} are based on the data for
89 galaxies compiled by \citet[][arXiv eprint version, updated on 21/02/2018]{Beck2013}.
If a galaxy in their Table~5 contained any indication of magnetic field parallel
to the galactic disc (as specified in the `Structure' column), it was counted as a galaxy
containing a large-scale magnetic field.
{Those galaxies listed \textit{only} as X-shaped, vertical, perpendicular to the plane 
or exhibiting halo spurs, as well as those listed as having no ordered field, 
were counted as not containing any large-scale magnetic field.}
The galactic stellar masses were obtained from the S4G catalogue \citep{S4G}, except for
M31 \citep{Chemin2009}, LMC \citep{Kim1998}, NGC~4236 \citep{Pezzulli2015}, 
NGC~4945, IC~342 \citep{Sofue2016}, NGC~7331, NGC~2403 and NGC~6946 \citep{deBlok2008}.
The data are consistent with a decrease in the fraction of galaxies that host significant
large-scale magnetic fields for masses below $10^{10}\msun$ but the number of such galaxies 
is too small (8 and 11 galaxies for the two lowest mass ranges) to reach any confident 
conclusions.
Apart from the largest mass range, both galaxy formation models predict a smaller fraction of
galaxies with large-scale magnetic fields than that indicated by observations.
It is possible that null detections are under-reported,
or that there is selection bias (e.g., galaxies being targeted because they 
are bright in the radio range), and such biases would tend to enhance the 
reported detection fraction. 
On the other hand, it is possible that in some galaxies the large-scale magnetic field
was too weak to be detected, yet still above the $0.05\muG$ cut-off, 
which would tend to reduce the detection fraction.
Another reason for the discrepancy could be an inaccurate estimation of the galactic stellar disc and bulge
masses in the galaxy formation models. These issues will be addressed elsewhere.
Nevertheless, it is encouraging that theory and observation agree that a significant fraction 
of galaxies at $z=0$ do \textit{not} contain large-scale magnetic fields,
and that this fraction appears to be higher for dwarf galaxies than for MW-like galaxies.

\begin{figure*}
 \centering
 \includegraphics[width=0.24\textwidth]{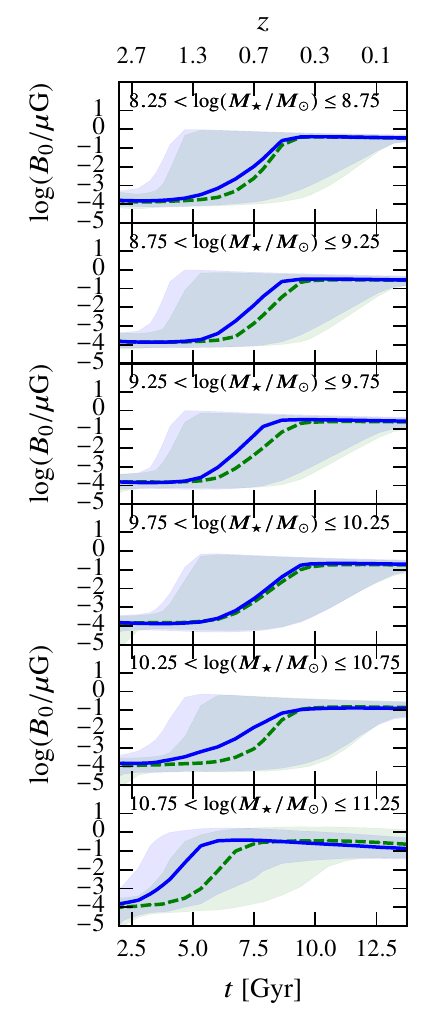}\hfill
 \includegraphics[width=0.24\textwidth]{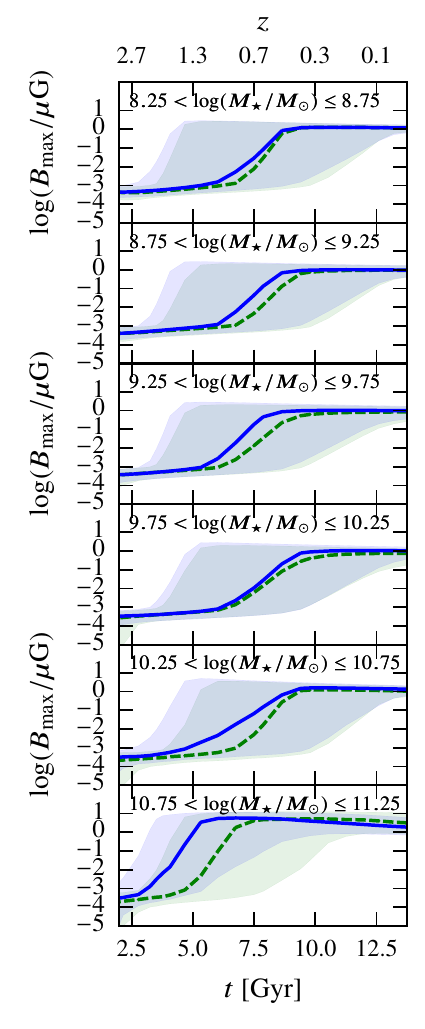}\hfill
 \includegraphics[width=0.24\textwidth]{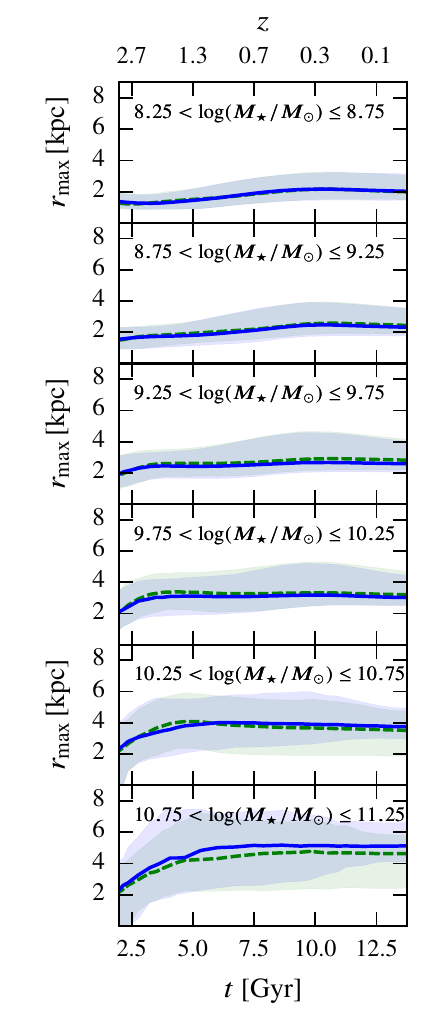}\hfill
 \includegraphics[width=0.24\textwidth]{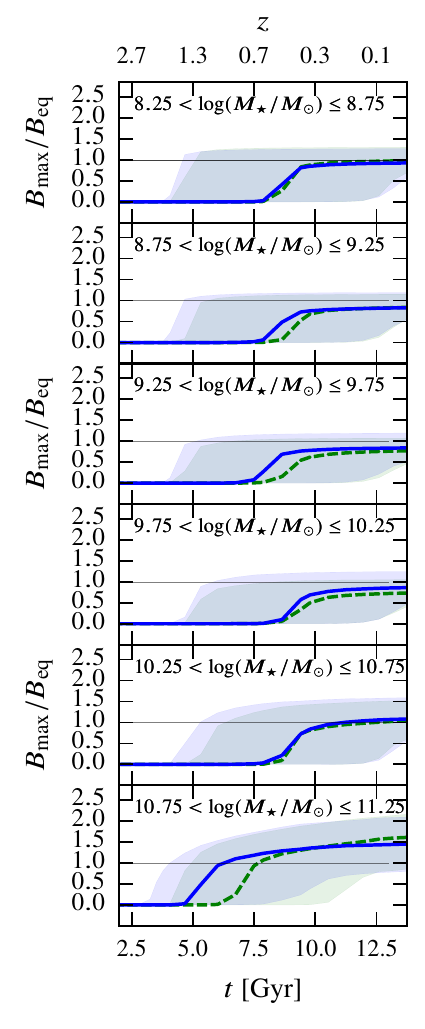}\\
 \includegraphics[width=0.24\textwidth]{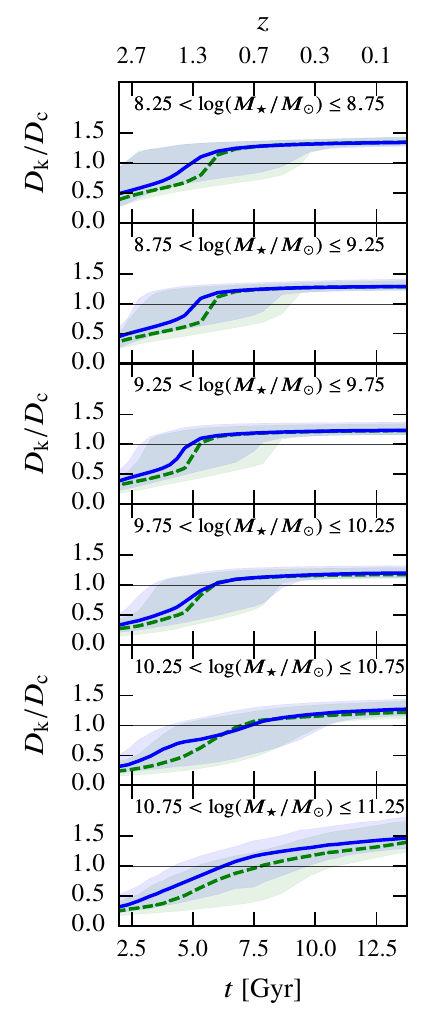}\hfill
 \includegraphics[width=0.24\textwidth]{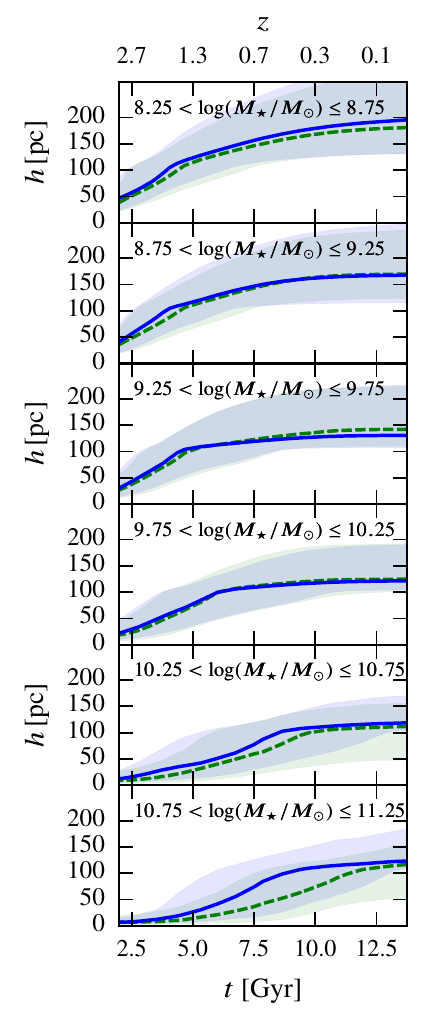}\hfill
 \includegraphics[width=0.24\textwidth]{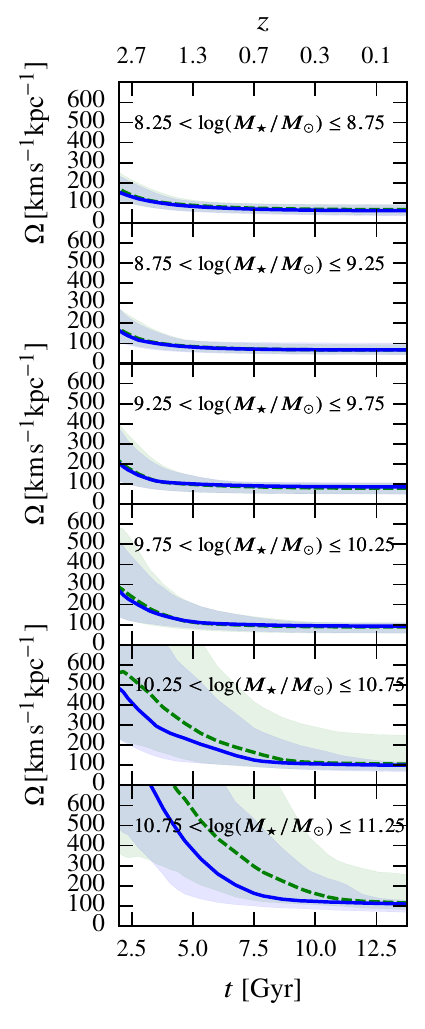}\hfill
  \includegraphics[width=0.246\textwidth]{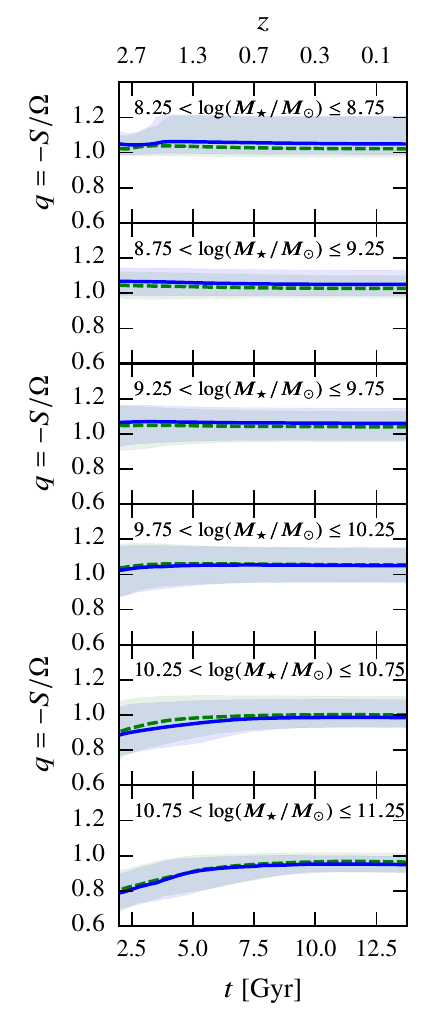}
\caption{\label{fig:theorists_Bgrid}
Evolution of (left to right and top to bottom): 
(i)~the rms large-scale magnetic field strength ${B}_0$ of equation~\eqref{eq:defBeavg},
(ii)~the maximum large-scale field strength ${B}\ma$ defined in
	equation~\eqref{eq:defBmax},
(iii)~galactocentric radius $r\ma$ where $\mean{B}(r\ma)={B}\ma$,
and the following quantities at $r=r\ma$:
(iv)~the ratio ${B}\ma(r\ma)/B\eq$, where $B\eq$ corresponds to energy equipartition
	with interstellar turbulence, equation~\eqref{Beq},
(v)~the ratio of the kinematic dynamo number $D\kin(r\ma)$ of equation~\eqref{Dkin} to its 
	critical value $D\crit(r\ma)$ of equation~\eqref{eq:Dc} 
(vi)~the scale height of the diffuse gas $h(r\ma)$,
(vii)~the angular velocity $\Omega(r\ma)$, and 
(viii)~dimensionless rotational shear rate; $q=1$ for a flat rotation
curve and $q=0$ for a solid-body rotation.
Only those spiral galaxies are included that have ${B}\ma>0.05\muG$ at $z=0$ in
each stellar mass interval indicated in the legend of each frame. The solid (blue) and
dashed (green) curves show the median of the distribution for the \Lacey and \Gonzalez
galaxy formation models, respectively, and the shaded areas of matching colour correspond
to the interval between the 15th and 85th percentiles.
        }
\end{figure*}

\begin{figure}
 \centering
 \includegraphics[width=0.3\textwidth]{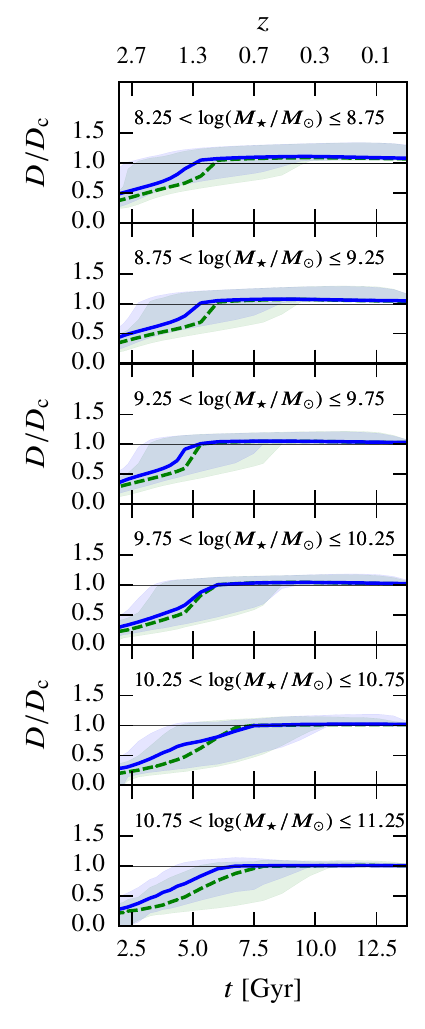}
 \caption{
 As in Fig.~\ref{fig:theorists_Bgrid} but showing
 \label{fig:Dgen}the ratio of the dynamo number affected by magnetic helicity, 
 $D(r\ma)$ of \eqref{D}, to $D\crit(r\ma)$.
 }
\end{figure}

\subsection{Magnetic history of galaxies}
\label{sec:origin}

In this section we consider in detail the population of galaxies that possess a 
large-scale magnetic field \red{at $z=0$} to understand when it was amplified to the present-day strength.
Figure~\ref{fig:theorists_Bgrid} illustrates the time evolution of the 
rms and maximum strengths of the mean galactic magnetic
fields as well as the derived parameters of the gaseous discs 
and dynamo number in various mass intervals. Spiral galaxies with
${B}\ma>0.05\muG$ in
six stellar mass ranges were selected at $z=0$, and their evolution was traced back to
the beginning of the \glf simulation. For most galaxies, the large-scale magnetic field
was amplified to the present-day level between $z=1$ and $z=3$. 
In each mass interval, there is a more or less well-defined epoch, of 
$\delta z\simeq1$ in duration, when the magnetic field starts growing 
and galaxies reach a steady dynamo state after about $8.5\Gyr$ of evolution.
The typical radius where the mean magnetic field strength is maximum varies little with time,
ranging from $1\lesssim r/\kpc\lesssim 2$ in the smallest stellar mass range, to
$2\lesssim r/\kpc\lesssim 7$ for the most massive galaxies.
This radius is usually close to that of maximum rotation shear near the turnover 
radius of a flat rotation curve.
The panel showing ${B}\ma/\Beq$ indicates that the most massive galaxies tend to
have large-scale magnetic fields above equipartition for $z<0.7$, while most
galaxies of a smaller stellar mass never quite reach $\Beq$.

The choice of a specific SAMGF has little impact except for the largest mass bin, where, in the
\Lacey model, many galaxies produce large-scale magnetic fields approximately
$2\Gyr$ earlier than in \Gonzalez.
{The main reason for this difference 
is the larger rotation rates in the former model.}

Given the critical value of the dynamo number $D\crit$ (equation~\ref{eq:Dc}), 
the mean-field dynamo activity is controlled by the diffuse gas scale height, $h$, 
angular velocity, $\Omega$, and shearing rate, $S=r\partial\Omega/\partial r$,
as $D\propto h^2\Omega S$. 
The time evolution of each of these quantities at the
radius of maximum field strength is shown in Fig.~\ref{fig:theorists_Bgrid}.
It does not seem that the control of the dynamo activity can be attributed to a
single dominant variable.
The angular velocity of galactic rotation first decreases with time and then stagnates
(as galaxy sizes grow faster than circular velocities
perhaps because of the difference in the rates of dark matter and gas accumulation)
while the shearing rate $S$ follows the behaviour of $\Omega$
(the ratio $q=-S/\Omega$ at $r=r\ma$ is approximately constant in time
as shown in the lower-right panel of Fig.~\ref{fig:theorists_Bgrid}).
Both the reduction in the rate of decrease of $\Omega$ (and $|S|$) 
and the steady increase of \red{$h$} play a role in the establishment of a supercritical dynamo.
This seems to apply to all stellar mass bins and redshifts,
though the relative importance of these two factors varies with stellar mass.
We also note that the rapid rise of $D/D\crit$ 
and of ${B}_0$ occurs earlier for galaxies with higher stellar mass.
Moreover, the larger the stellar mass,
the less scatter in the time at which this rapid growth phase happens.

Finally, comparing $D\kin/D\crit$ with $D/D\crit$ (the first panel in
the bottom row of Fig.~\ref{fig:theorists_Bgrid} and Fig.~\ref{fig:Dgen}) allows one to
appreciate the magnetic contribution to the $\alpha$-effect
(the term $\alp\magn $ in equations~\ref{dBrdt}--\ref{dalpha_mdt})
and the associated quenching of the mean-field dynamo.
As expected, the dynamo number modified by the magnetic field gradually 
approaches the critical value 
as magnetic field grows, so that $D/D\crit\approx1$ whenever $D\kin/D\crit>1$.

\begin{figure*}
 \centering
 \includegraphics{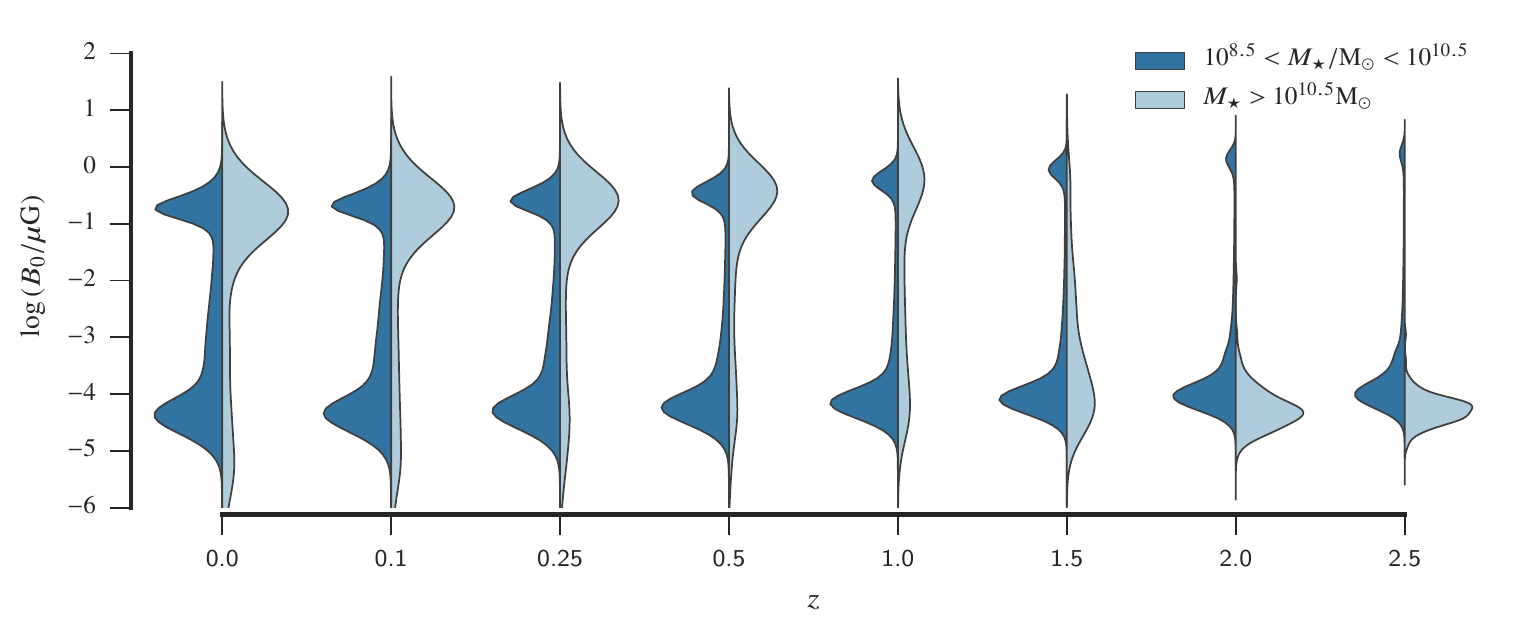}\vspace{-0.7em}
\mbox{}\hspace*{-0.7em}
\includegraphics{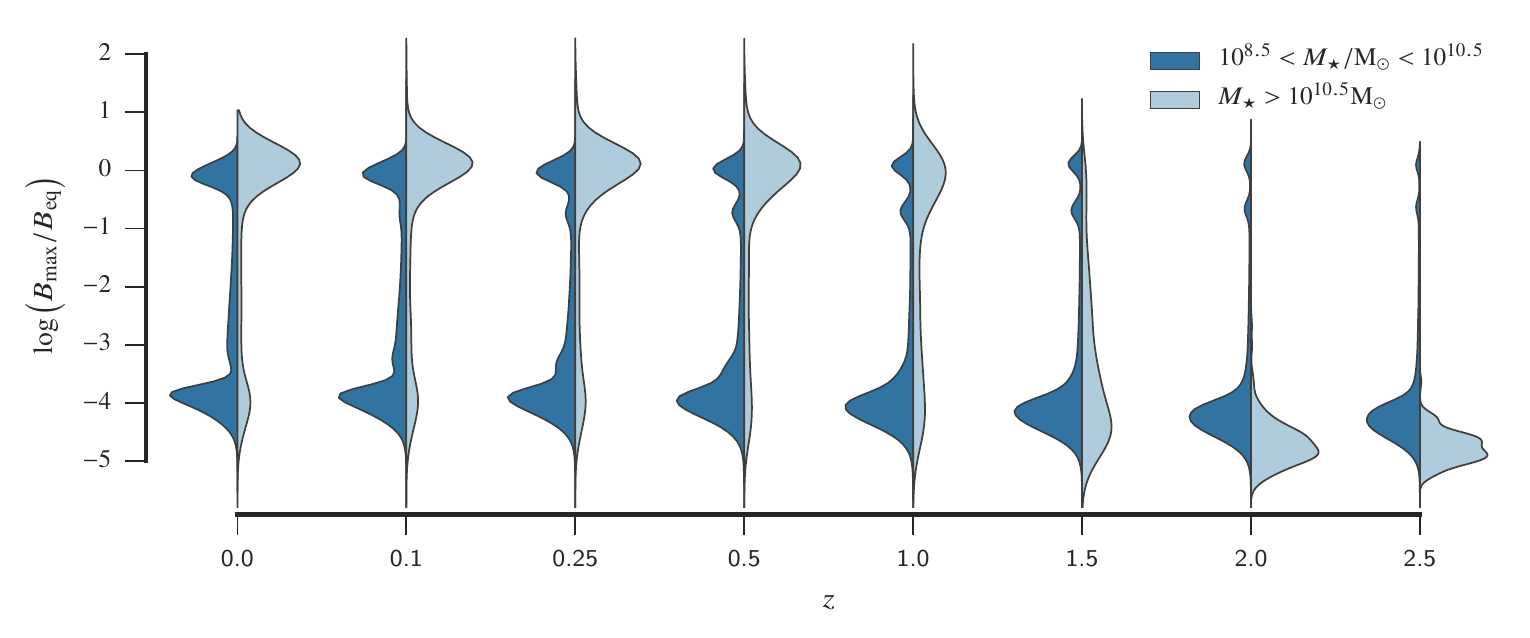}\vspace{-0.7em}
 \includegraphics{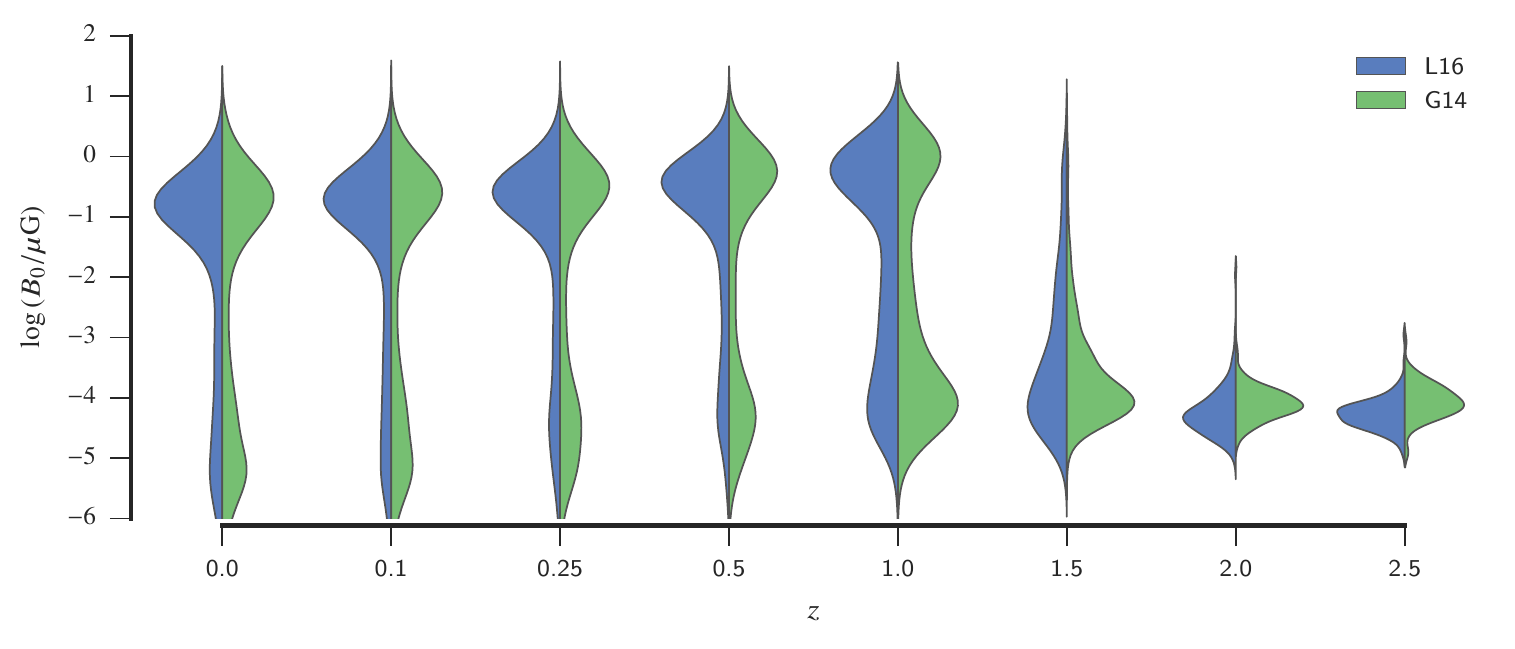}\vspace{-0.7em}
 \caption{\label{fig:violin}
The variation with redshift (the horizontal axis) of the probability density of the
large-scale magnetic field strength
(the rms field in the upper and lower panels, and the ratio of the maximum to the
equipartition field in the middle panel)
of spiral galaxies separated at each redshift into two stellar mass bins,
$8.5<\log(M_\star/\msun)<10.5$ (dark blue) and $\log(M_\star/\msun)>10.5$ (light blue).
The top two panels are obtained for the \Lacey model,
while the bottom panel illustrates the difference between the \Lacey (blue) and \Gonzalez 
(green) models for the massive galaxies, $M_\star>10^{10.5}\msun$, in terms of the rms
magnetic field strength.
    }
\end{figure*}
\begin{figure*}
 \centering
 \includegraphics{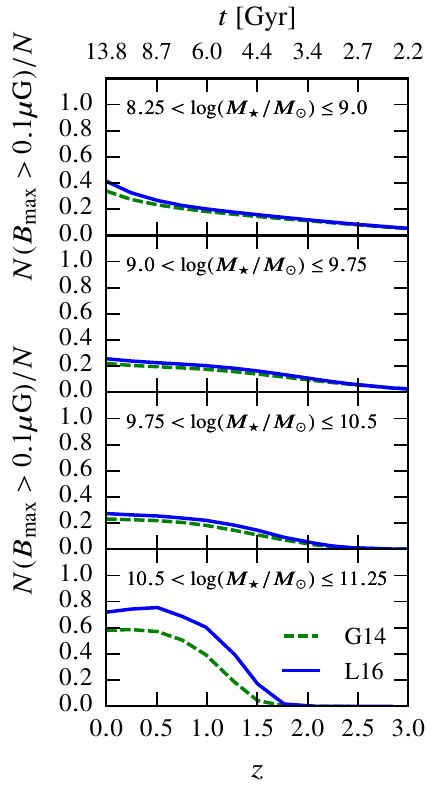}
 \includegraphics{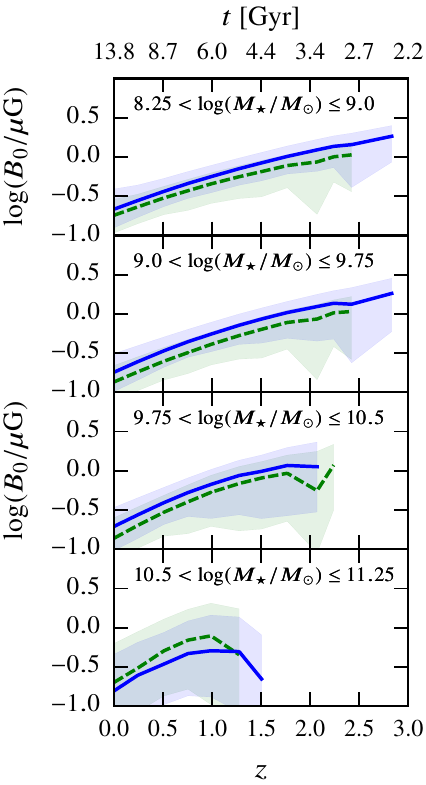}
 \includegraphics{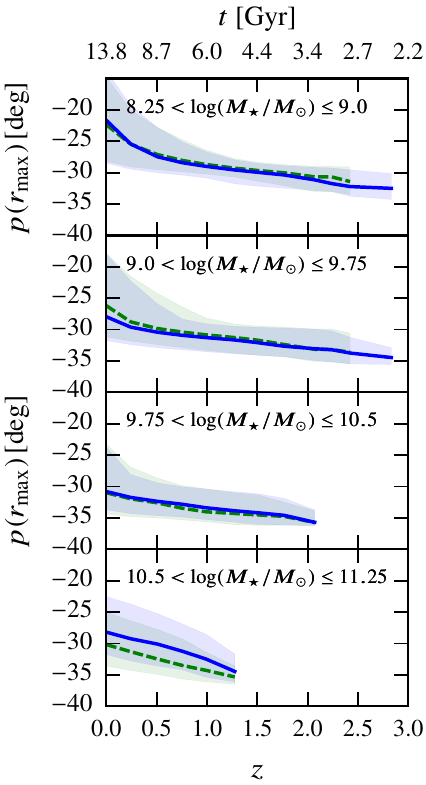}
 \includegraphics{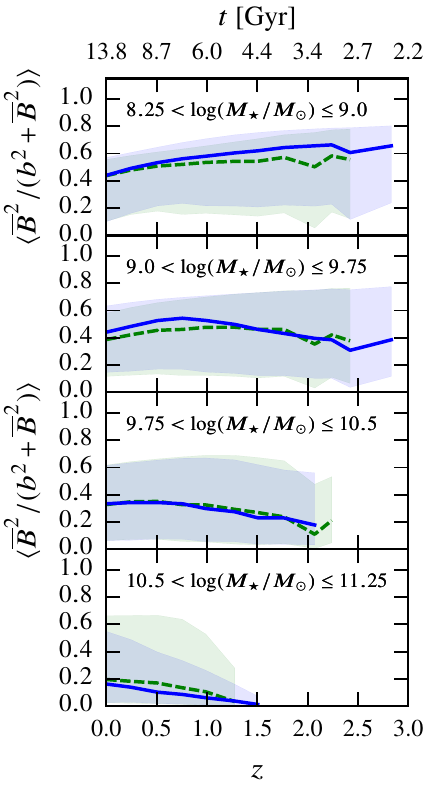}
 \caption{ \label{fig:observers_plots}
The evolution of magnetic field parameters for galaxies in the mass intervals shown in the legends  (from left to right): 
(i)~the fraction of spiral galaxies with ${B}\ma>0.1\muG$;
(ii)~the rms large-scale field strength ${B}_0$;
(iii)~magnetic pitch angle at the radius of maximum field, $p(r\ma)$; and 
(iv)~the ratio of the large-scale to total magnetic field strengths
averaged over the galactic disc.
The results are shown only for spiral galaxies with significant
(exceeding $0.05\muG$) large-scale magnetic
fields selected by mass at each redshift. 
The solid (blue) and dashed (green) curves show the median 
of the distribution for the \Lacey and \Gonzalez models, respectively. The shaded areas 
correspond to the interval between the 15th and 85th percentiles.
    }
\vspace{-1em}
\includegraphics[width=\textwidth]{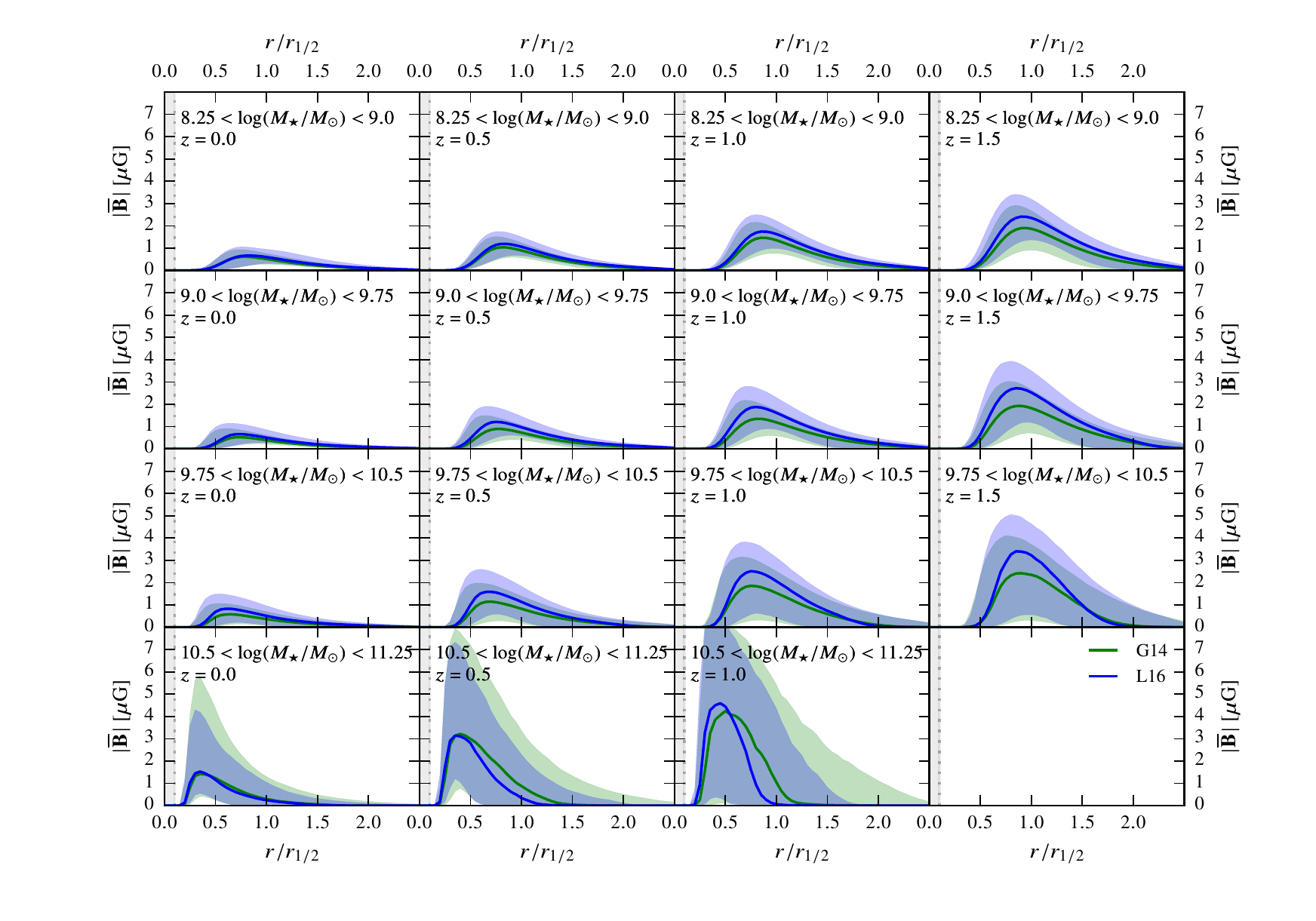}\vspace{-4.2em}
 \caption{Radial profiles of large-scale magnetic field strength,
 for different choices of redshift and stellar mass (as indicated in the legends) for 
 galaxies with ${B}\ma>0.1\muG$.
Solid curves represent the median values and shaded areas show the 15th and 85th
percentiles.
          \label{fig:profiles_B}}
\end{figure*}

\begin{figure*}
 \centering
 \includegraphics[width=\textwidth]{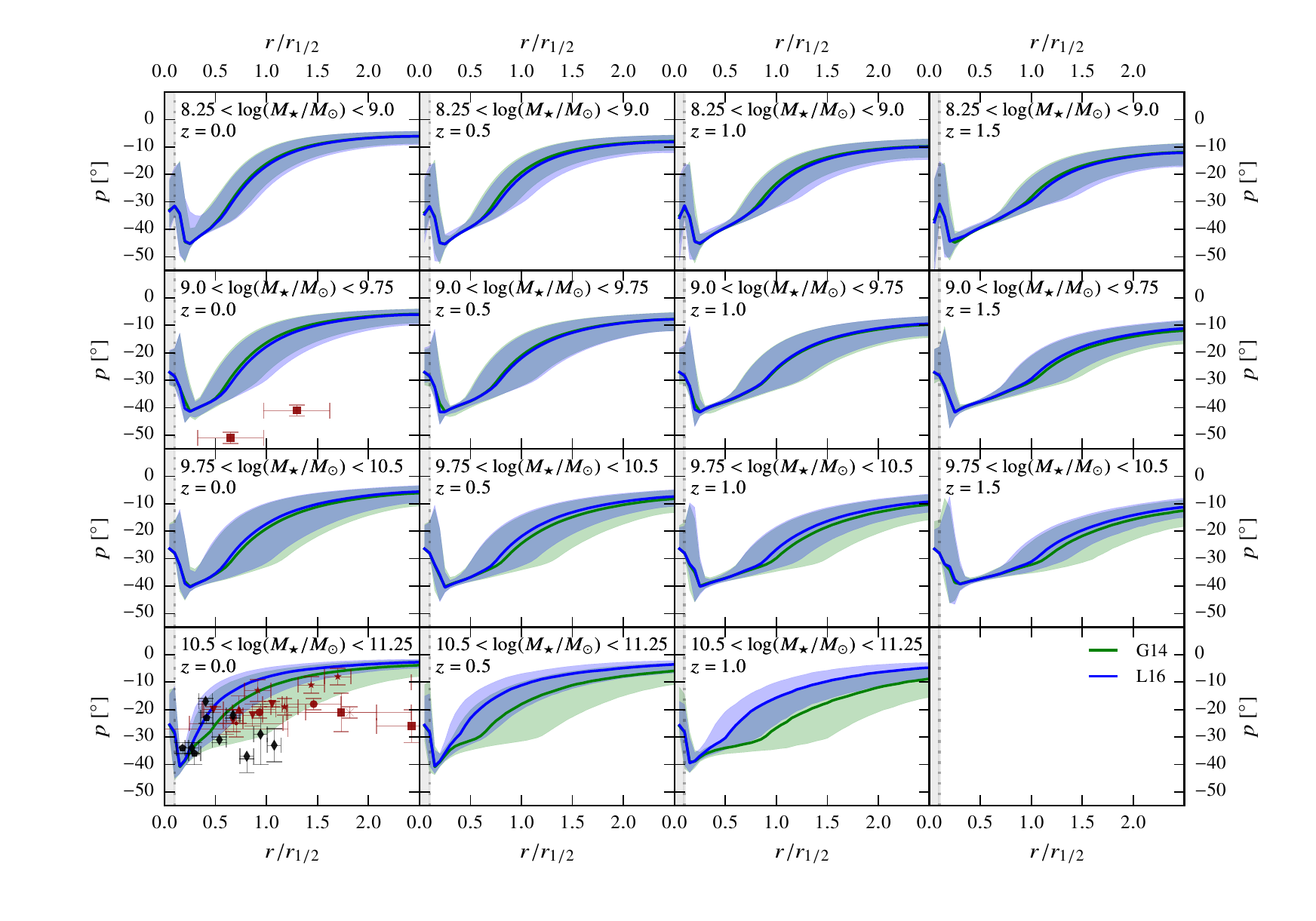}\vspace{-3em}
 \caption{Pitch angle of the large-scale magnetic field $p=\arctan{\mean{B}_r/\mean{B}_\phi}$
versus galactocentric radius for different choices of redshift and stellar mass 
(as indicated in the panel legends) for galaxies with 
{${B}\ma>0.1\muG$}.
Solid curves show median values and shaded areas span the range between 
the 15th and 85th percentiles (blue for \Lacey and green for \Gonzalez). 
Points with error bars at $z=0$ show observational estimates for M31 , M33, M51, M81, 
NGC~1566, NGC~253, NGC~6946, IC~342 (red) and barred galaxies NGC~1097 and NGC~1365 (black)
\citep[see][for references]{Chamandy+16}; horizontal error bars represent the radial range
to which the estimates belong.
           \label{fig:profiles_p}}
\end{figure*}

\begin{figure*}
 \centering
 \includegraphics[width=\textwidth]{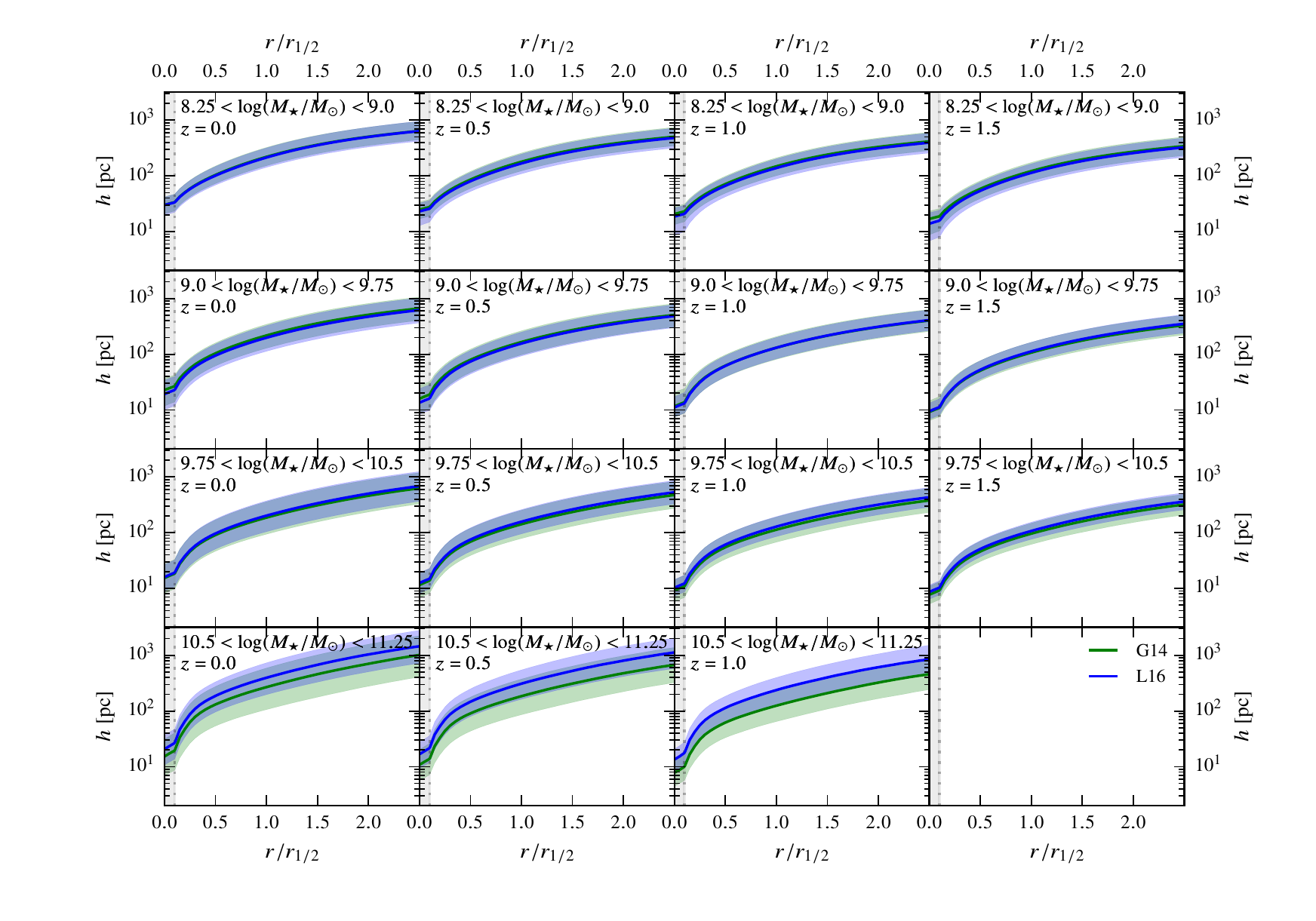}\vspace{-3em}
 \caption{As in Figs~\ref{fig:profiles_B} and \ref{fig:profiles_p} but for the
 diffuse gas scale height $h$.
          \label{fig:profiles_h}}
\end{figure*}

\subsection{Cosmological evolution of galactic magnetism}\label{CEoGM}
The previous section is focused on the evolution of the
population of galaxies with significant large-scale magnetic field at redshift $z=0$.
These are the galaxies that would be selected for magnetic field studies in the
nearby Universe. The sample is obviously biased. To provide a more complete picture,
here we consider \red{constructing samples of galaxies from our simulation using}
criteria characteristic of
observations of high-redshift galaxies, e.g., the surface brightness.
Since most galactic observables are related to the galactic stellar mass, 
we use this variable as a proxy for any other observable.

Figure~\ref{fig:violin} shows how the probability density distributions of 
the rms and maximum field strengths of the large-scale magnetic field evolve
for $0<z<2.5$ for two galactic mass ranges, $8.5<\log(M_\star/\msun)<10.5$ and $\log(M_\star/\msun)>10.5$.
At each redshift, a significant fraction of the galaxies have weak magnetic field 
strengths of order $10^{-4}\muG$. 
The fraction of galaxies which do contain significant large-scale magnetic field
\emph{decreases} with redshift (and increases with time) in agreement with 
\citet{Paper1}. This suggests that the assumption of \citet{Paper1}, relaxed in this work, 
that the mean-field dynamo is approximately in a steady 
state in each galaxy at any given redshift, is fairly robust.
This is consistent with the analysis of Section~\ref{sec:origin},
where it is shown that once $D/D\crit$ has reached unity, 
it does not deviate very much from that value.
If it was common for galaxies to deviate from the steady state once they 
produced significant large-scale fields, 
then this would have resulted in more scatter about $D/D\crit=1$ 
than what is seen in Fig.~\ref{fig:Dgen}.
The middle panel of Fig.~\ref{fig:violin} shows the ratio of the maximum field
strength to the equipartition field strength -- Eq.~\eqref{Beq} --,
the typical values $B\ma/B\eq(r\ma)$ for the galaxies with significant
large-scale fields remain approximately constant, which is expected if the galaxies had
reached steady-state.

The top panel of Fig.~\ref{fig:violin}
shows that galaxies with stellar masses $\log(M_\star/\msun)>10.5$ undergo 
a transition from being devoid of significant large-scale fields at $z=2$ 
to mostly having significant large-scale fields at $z=1$.
Galaxies with $\log(M_\star/\msun)<10.5$, on the other hand, have a very different history.
At $z=2.5$, there is already a small but significant fraction of galaxies that 
contain large-scale fields of order a few $\!\muG$.
The fraction of galaxies that have significant large-scale fields 
(the ratio of areas around the local maxima in the probability density) increases
steadily with time,
whereas for the higher stellar mass population, it increases until $z\simeq0.5$,
after which it remains roughly constant (and even decreases slightly).
Thus, the overall evolution tends to be faster for the higher stellar mass population.
Furthermore, of those galaxies that host significant large-scale magnetic fields in the lower stellar mass population, 
the median value is larger than for the more massive galaxies,
while the width of the distribution is much narrower.

In the bottom panel of Fig~\ref{fig:violin}, we compare
the \Lacey and \Gonzalez models for $\log(M_\star/\msun)>10.5$. 
The transition redshift after which most of the massive galaxies contain significant
large-scale magnetic fields is lower in \Gonzalez than in \Lacey.
The results for small-mass
galaxies are indistinguishable for the two models and have, thus, been omitted.

We found in Section~\ref{sec:origin} that in galaxies that host a large-scale magnetic 
field, the typical field strength decreases with time in each mass interval considered.
This is further illustrated in the second column of Fig.~\ref{fig:observers_plots} where the
galaxies with negligible large-scale magnetic field were removed.
Thus, a survey of galaxies selected by mass (or a related quantity) at each redshift would
find a strong \emph{increase} of the typical field strength observed with redshift.
However, at higher redshifts, not only does the abundance of galaxies with a certain
stellar mass decrease (due to the evolution of the stellar mass function -- see,
e.g., Fig.~24 in \citealt{L16}) but so does the fraction of those which contain 
significant large-scale magnetic fields (the first column of Fig.~\ref{fig:observers_plots}).

The third panel of Fig.~\ref{fig:observers_plots} shows the evolution of the
magnetic pitch angle, given by $\tan p=\mean{B}_r/\mean{B}_\phi$ 
(with $-90^\circ<p\le90^\circ$), that characterises how tightly wound is
the large-scale magnetic field spiral \citep[see][and references therein]{Chamandy+Taylor15,Chamandy+16}.
The pitch angle, reported at the radius of maximum magnetic field strength, generally
ranges from $-20\deg$ to $-30\deg$ (with negative values corresponding to a trailing spiral). 
The typical value of $p(r\ma)$ depends on the galaxy mass, 
and, in all cases, the median value of $p(r\ma)$ decreases \red{(becomes more negative)}
with redshift: magnetic lines become more tightly wound as the galaxy evolves.

The rightmost panel of Fig.~\ref{fig:observers_plots} shows the evolution of the 
degree of order in the magnetic field defined as the
ratio of the mean to the total magnetic field averaged over the disc surface, 
$P=\langle \mean{B}^2/(\mean{B}^2+b\turb^2)\rangle$.
This quantity varies significantly with both galactic stellar mass and redshift:
massive galaxies have less ordered magnetic fields, $0<P\lesssim0.6$ 
with a median of $0.2$ at $z=0$, while the smallest galaxies, 
$8.25<\log(M_\star/\msun)<9.0$, have $0.2\lesssim P\lesssim0.8$ with a median
of about $0.45$ at $z=0$. Since $P$ is related to the fractional polarisation of synchrotron
emission (that can be further reduced by Faraday effects), these results indicate that the 
fractional polarisation is expected to decrease strongly with redshift if only massive 
galaxies are selected at each redshift. The opposite trend is expected for galaxies of
the lowest stellar mass.

\subsection{Radial magnetic profiles}\label{RMP}

Figures~\ref{fig:profiles_B} and~\ref{fig:profiles_p} present the radial profiles of
the large-scale magnetic field strength and magnetic pitch angle, respectively.
At redshifts $z=0$, $0.5$, $1.0$ and $1.5$, galaxies were selected by
mass -- in the four intervals shown -- and, to account for variations in the sizes of
galactic discs, the galactocentric radius is normalised to the half-mass radius of 
each galaxy, $r_{1/2}$.
The grey shaded area along the left-hand vertical axis in each panel indicates the 
region where the angular velocity 
is truncated as described in Section~\ref{sec:angular_velocity}.
Our model underestimates magnetic field strength in this region
because of this truncation; furthermore, the thin-disc and, possibly, $\alpha\omega$
approximations are not justified there \citep{Ruzmaikin+88,Chamandy16}.
Large-scale magnetic fields are expected to be strong
in the central parts of galaxies.

The half-mass radii of individual nearby galaxies were estimated as
$r_{1/2}=0.4 R_{25}$, where $R_{25}$ is the radius of the 
$25\,\text{mag\,arcsec}^{-2}$ isophote from the LEDA database, and the stellar mass
was obtained from the S4G \citep{S4G} catalogue. We have calculated 
$r_{1/2}/R_{25}$ for 101 galaxies in the 2MASS catalogue to obtain
the median and mean values of 0.39 and 0.44, respectively, with the sample
standard deviation of 0.16. 

Figure \ref{fig:profiles_B} shows that the relative radius $r\ma/r_{1/2}$
where magnetic field strength is maximum is smaller at lower redshifts 
in each mass interval, apparently because the turnover radius of galactic 
rotation curves, where the rotational shear is maximum, decreases with time.
There is generally good agreement between the two galaxy formation models, \Lacey and 
\Gonzalez. 
Galaxies with smaller stellar mass have more extended profiles in the unit of half-mass
radius,  showing that the magnetic field profile does not scale linearly with the
galaxy size. 
Large-scale magnetic fields are first amplified closer to the galactic centre,
where the angular velocity and its shear are larger, and then spread outwards in the form
of a magnetic front 
\citep{Moss+Sokoloff98,WSSS04}. Thus, our results suggest that
the rate of growth of galactic discs is larger than the rate of spread of the magnetic field.
Consistently with Figs.~\ref{fig:PDFs} and \ref{fig:violin}, the range of field
strengths depends on galaxy stellar mass, being much wider at larger masses.

In the $z=0$ (leftmost) column of Fig.~\ref{fig:profiles_p}, we show observational 
estimates of the magnetic pitch angles in nearby
galaxies compiled by \citet{VanEck2015} and \mbox{\citet{Chamandy+16}}.
These data points are supplied with horizontal error bars that  
show the range of radii where the value was reported by \citet{VanEck2015}. 
(We note that our estimates of the half-mass radius for nearby galaxies may not be
reliable even if they are statistically accurate; this may have affected the
observational data points in Fig.~\ref{fig:profiles_p}.)
Magnetic pitch angle, shown in Fig.~\ref{fig:profiles_p}, decrease in magnitude  with 
radius, an overall trend consistent with observations 
\citep[Section~4.6 in][]{Beck15}.
The two galaxy formation models used lead to significantly different predictions 
for the pitch angle profiles of the most massive galaxies, with \Gonzalez producing larger 
$|p|$.
At $r/r_{1/2}<1.5$, predictions of  the \Gonzalez model are marginally consistent with
the observations of nearby galaxies while the \Lacey model predicts somewhat too small
values of $|p|$.
{
Better agreement would require more accurate estimates of parameters like $l_0$
and $v_0$, which are likely to vary within and between galaxies.
In addition, 
other} refinements of 
{our} dynamo model
may improve the agreement, {such as accounting for} 
the effects of spiral patterns, galactic outflows and accretion flows 
\citep[see][for a discussion]{Chamandy+16}.

The scale height of diffuse gas as shown in Fig.~\ref{fig:profiles_h} increases 
with galactocentric radius -- the gas disc is flared. At $r/r_{1/2}\gtrsim0.5$, 
the scale height increases nearly
exponentially, mainly because of the exponential radial decrease in the stellar gravity
field. The exponential gas disc flaring agrees with \ion{H}{i} observations in the MW 
\citep[][and references therein]{KK09} and other galaxies. \red{
\citep[see e.g.][]{Peters2017,OBrien2010},
and lends support to the ISM model in evolving galaxies developed in this paper.
Also \citealt{Chamandy+16} shows that flaring scale height profiles are
necessary in order to explain the observed pitch angles of nearby
galaxies using a general mean field dynamo model}.

\section{Discussion}\label{Disc}

The seed magnetic field for the mean-field dynamo, estimated as 
$\widetilde{B}\simeq5\times10^{-4}\muG$ in Section~\ref{sec:floor}, is provided by the 
fluctuation dynamo action of interstellar turbulence. This mechanism to generate random magnetic 
fields can be active continuously as soon as star formation in an evolving galaxy becomes intense 
enough to 
drive pervasive random flows in the interstellar gas. Unlike primordial seed magnetic fields, 
this mechanism constantly seeds galactic mean-field dynamos and therefore can support 
re-launching of mean-field dynamo action in galaxies that became temporarily incapable of 
sustaining mean-field 
dynamo action in the course of their evolution, e.g., due to the destructive effects of mergers on 
the galactic rotation and gaseous disc.

The presence of random magnetic fields at all stages of evolution of star-forming galaxies has 
an important implication for the confinement of cosmic rays. The Larmor radius of a 1\,GeV 
relativistic proton is smaller than the characteristic disc scale height of $200\pc$ in a magnetic field 
as weak as $10^{-5}\muG$. The fluctuation dynamo generates random magnetic fields of a strength 
comparable to that corresponding to equipartition with the turbulent kinetic energy density,
a few microgauss. Therefore, cosmic rays can be efficiently confined to galaxies from the 
earliest stages of their evolution \citep{Zw13}.
This justifies our inclusion of cosmic ray 
pressure in the calculation of the gas scale height in hydrostatic equilibrium in 
Appendix~\ref{ap:Pgas} and equation~\eqref{eq:Pleft}.

As Fig.~\ref{fig:violin}} shows, the evolution of galactic large-scale
magnetic fields is rather non-trivial: on the one hand, the fraction of galaxies that 
host such a field increases with time but, on the other hand, the field strength decreases.
This has interesting implications for the effects of magnetic field on the stellar feedback
on galactic evolution via galactic outflows. As discussed by
\citet{EGSFB18}, large-scale
magnetic fields produced by dynamo action are effective in quenching galactic outflows, with 
the dependence of the outflow speed on magnetic field strength fitted by
\begin{equation}\label{uzB}
V_z\propto \left[1+0.5 (\mean{B}/B\eq)^n\right]^{-1}\,,
\end{equation}
with $n$ in the range 3--4 \citep[see also][]{BGE15,EGSFB18proceedings}. The dependence on $\mean{B}$ is strong,
implying a very efficient suppression of an outflow. Simulations of the supernova-driven ISM,
on which this result relies, do not yet contain cosmic rays; their pressure may make the 
effect weaker. 
Nevertheless, the possibility of a strong effect of the galactic mean-field dynamo action
on the stellar feedback needs to be considered carefully. Taken at their face value, our
results suggest that the outflows can be more readily suppressed at larger redshifts where magnetic
field strength is larger. This question requires further analysis.

In every mass range (perhaps except for the most massive galaxies -- see 
Fig.~\ref{fig:fractions}), there are galaxies that 
do not have any significant large-scale magnetic fields. Apart from the statistical scatter in the
magnitude of rotation speed and its shear (resulting from the random nature of the evolution of
the galactic angular momentum), reflected in the scatter of the kinematic
dynamo number $D\kin$, galaxy mergers can contribute to disturbing the mean-field dynamo action.
In the model that we use, the major mergers are assumed to destroy galactic gaseous discs 
and disperse magnetic fields, and then the dynamo may be launched again. It is not surprising
then that massive galaxies have a wider range of field strengths as they 
have more complex formation history than galaxies of smaller mass.

\red{
The present work focused on the magnetic fields in galactic discs.
Galactic discs are much brighter in synchrotron emission than the haloes \citep{Wiegert2015}
and therefore dominate in such observables as the galactic radio luminosity function.
On the other hand, the magnetic field in galactic haloes --
which was not considered here -- may be significant for
Faraday rotation measure studies
and in detailed radio characterisation of
resolved edge-on galaxies, and \luke{has} been explored in recent works
\citep[see e.g.][]{Henriksen2018}. 
A treatment of the magnetic field in haloes which consistently accounts for the
galaxy evolution requires detailed understanding of the magnetised winds produced
by galaxies \citep[see][and references therein]{Heesen2018}
and is \luke{the} subject of our present active research.
}

\section{Summary and conclusions}\label{SaC}
We have developed an ISM model for a large sample of more than a million evolving 
galaxies based on SAMGF and used it to explore statistical properties of galactic
dynamos and magnetic fields that they produce. The ISM model includes those 
ingredients that control fluctuation and mean-field turbulent dynamos, galactic rotation 
curves and flared gas discs in particular (Figs.~\ref{fig:theorists_Bgrid} and 
\ref{fig:profiles_h}). Seed magnetic fields in our model of the galactic mean-field dynamo
are continuously replenished by fluctuation dynamo action, which allows the mean-field
dynamo to be reactivated after major galactic mergers or other events destructive for 
galactic discs and their magnetic fields. (This mechanism of galactic 
seed magnetic field generation is therefore more flexible than primordial mechanisms.) Such 
disruptions are especially common in the formation history of massive galaxies. As a result,
the more massive is a galaxy, the later, on average, it becomes able to support a
sustained mean-field dynamo action and a microgauss-strength large-scale magnetic field
(Fig.~\ref{fig:observers_plots}).

A significant fraction (dependent on galactic mass) of galaxies possess no large-scale
magnetic fields, with a minimum of about 20 per cent for the stellar mass 
$M_\star\simeq10^{10}\msun$ (Figs.~\ref{fig:fractions} and \ref{fig:violin}),
some galaxies cannot support the mean-field dynamo (Fig.~\ref{fig:PDFsActPas}) because
their rotation rate and/or velocity shear is too weak or the gas disc is too thin.

Our results suggest that large-scale magnetic fields can be amplified to a microgauss 
strength in $2\text{--}3\Gyr$, starting at $z=3\text{--}2$ depending on the galactic mass
(Fig.~\ref{fig:theorists_Bgrid}).
We stress that this time scale is much shorter than the often quoted estimate of $10\Gyr$ (or 
more) based on estimates only applicable in the Solar neighbourhood of the MW. The 
galactocentric radius where the large-scale magnetic field is maximum ranges from about
$2\kpc$ in low-mass galaxies to $5\kpc$ in massive ones.
In general, dynamo action starts later in more massive galaxies 
(Fig.~\ref{fig:observers_plots}) but this depends subtly on sample selection criteria: 
for galaxies selected at $z=0$ for their strong
magnetic field, the trend is reversed and more massive galaxies produce their magnetic 
fields earlier (Fig.~\ref{fig:theorists_Bgrid}).
The differences between magnetic histories of galaxies of various masses are largely
due to the differences in their assembly; our model assumes that each major merger
disperses galactic gas discs with their magnetic fields, and dynamo action resumes
again after such events.
This shows that interpretations of
either observations or simulations of galactic magnetic fields in high-redshift galaxies 
require extreme care and attention to sample biases and selection effects.

Another non-trivial aspect of our results is a prediction that the fraction of
galaxies with significant large-scale magnetic fields increases with time but the
strength of their magnetic fields decreases due to the depletion of interstellar gas as
a galaxy evolves (Figs.~\ref{fig:violin} and~\ref{fig:observers_plots}). 

The two versions of the \glf galaxy formation model considered (\Lacey and \Gonzalez) 
lead to broadly similar magnetic field forms and evolution patterns but there is some
variance too. The most prominent difference is that they predict different fractions of 
galaxies with strong large-scale magnetic fields (Fig.~\ref{fig:violin}). This and other 
differences can be used to constrain galaxy formation models using magnetic field 
observations in statistically representative samples of galaxies, either in the nearby
Universe or at high redshifts.

\section*{Acknowledgements}
\red{We thank the referee R.~N.~Henriksen for his comments. 
     We thank R.~Beck for useful suggestions.}
{LFSR thanks University of Cape Town for its warm hospitality during his visit.}
{LC gratefully acknowledges Newcastle University for its hospitality.}
AS and LFSR acknowledge financial support of STFC (ST/N000900/1, Project 2).
This work used the COSMA Data Centric system at Durham University, operated by the Institute for Computational Cosmology on behalf of the STFC DiRAC HPC Facility 
(\url{www.dirac.ac.uk}).
This equipment was funded by a BIS National E-infrastructure capital grant ST/K00042X/1, DiRAC Operations grant ST/K003267/1 and Durham University. DiRAC is part of the National E-Infrastructure.
This research has made use of NASA's Astrophysics Data System.

{\footnotesize
\noindent
\bibliographystyle{mnras}
\bibliography{refs}
}

\appendix

\section{Details of the galaxy modelling}
\subsection{Molecular gas fraction}
\label{ap:fmol}

To compute the molecular fraction of the interstellar gas density, we start 
by computing a \emph{notional} pressure
profile, $\widetilde{P}
(\Sigma_\star,\Sigma_\text{g})$, based on the approximate
expression for the mid-plane pressure suggested by \citet{Elmegreen1989},
\begin{equation}
\widetilde{P}(r) = \frac{\pi}{2} G \,\Sigma_\mathrm{g}(r) \left[\Sigma_\mathrm{g}(r)
                + \frac{v_0}{\sigma_\star(r)}
                \Sigma_\star(r)\right]\label{eq:Pelme}\,,
\end{equation}
where $\sigma_\star$ is the velocity dispersion of the stellar component, estimated
using
\begin{equation}
 \sigma_\star^2 = \pi G h_\star \Sigma_\star(r)\,,
\end{equation}
where the stellar scale height is $h_\star= f r_s$, with $f\approx0.137$
\citep{Kregel2002}.

Once $\widetilde{P}
(r)$ is known, the fraction of molecular gas can be computed as
\begin{equation}
    \kappa(r) =
\left[{\widetilde{P}(r)}/{P_0}\right]^\alpha\,,
\end{equation}
where $P_0=4.787\times10^{-12} \erg\cm^{-3}$ and
$\alpha=0.92$ were obtained from observations
\citep{BlitzRosolowsky2004,BlitzRosolowsky2006,Leroy2008}.

\subsection{Rotation curves}
\label{ap:rotation}

To be able to obtain an accurate solution for the magnetic field, 
good knowledge of the rotation curve is required. 
We reconstruct the rotation curve from \glf output in the following way. 
The disc is assumed to be thin and to have an exponential surface density profile,
which leads to the following rotation curve \citep{BinneyTremaine2008}:
\begin{equation}
 V_\text{disc}(y) \propto y\left[ I_0(y) K_0(y) - I_1(y)  K_1(y) \right]^{1/2}\,,
\end{equation}
where $y = r/(2r_s)$, and $I_n$ and $K_n$ are modified Bessel functions of the
first and second kinds, respectively, and $r_s$ is the disc scale radius.

We assume the mass in the bulge follows a \citet{Hernquist1990} profile
which leads to the following circular velocity profile
\begin{equation}
 V_\text{b} \propto \sqrt{\frac{r}{(r+r_\text{b})^2}}\,,
\end{equation}
where $r_\text{b}$ is a characteristic radius.

Finally the dark matter halo is assumed to follow an NFW profile, which leads
to a rotation curve \citep{Mo2010},
\begin{equation}
 V_{0}(r) = V_\text{vir} \left[
 \frac{1}{x} \frac{\ln(1+cx)-cx/(1+c)}{\ln(1+c)-c/(1+c)}\right]\,,
\end{equation}
where $x=r/r_\text{vir}$, $c$ is the halo concentration parameter,
$r_\text{vir}$ is the virial radius and $V_\text{vir}=V_0(r_\text{vir})$.

The initial dark matter profile is, then, corrected for the effect of the contraction
of the dark matter halo due to the presence of baryonic matter.
We use the assumption of adiabatic contraction, where individual mass shells of
the system are assumed to conserve mass and angular momentum after the contraction.
This is done finding, for each radius, $r$, the radius in the initial mass
distribution, $r_0$, such that
\begin{equation}
  r_0^2 V_0^2(r_0) = r^2\left[V_\text{disk}^2(r) +V_\text{b}^2(r)\right]
                        + (1-f_\text{b}) r_0^2 V_0^2(r_0)\,,
                        \label{eq:adiabatic_contraction}
\end{equation}
where $f_\text{b} = (M_\text{b}+M_\text{disk})/M_\text{total}$ is the baryon fraction.
From equation~\eqref{eq:adiabatic_contraction}, the final dark matter circular velocity
can be computed as
\begin{equation}
  V_\text{dm}(r) = V_0(r_0) \sqrt{1-f_\text{b}}\;.
\end{equation}

The various contributions can then be combined to obtain the total rotation curve
through
\begin{equation}
 V = \sqrt{V_\text{disk}^2+V_\text{b}^2+V_\text{dm}^2}\,.
\end{equation}

\newcommand{\rcyl}{r}

\subsection{Pressure profile}
\subsubsection{Hydrostatic equilibrium}
\label{ap:P}

Assuming that the vertical gradient of the total gas pressure balances gravity, we write
\begin{equation}
\deriv{P}{Z} = -\rho_\dd(\rcyl,Z) \deriv{\phi}{Z}\,,
\label{eq:apg}
\end{equation}
where $\rcyl$ is the cylindrical radial coordinate.
The gravitational potential satisfies the Poisson equation written here for a thin disc
\citep[e.g.,][p.~77]{BinneyTremaine2008}
\begin{equation}
{\deriv{^2 \phi}{Z^2}
}
= 4\pi G \sum_i\rho_i(Z)\,,\label{eq:poisson}
\end{equation}
where the sum includes the contributions from diffuse gas,
$\rho_\text{d}$; molecular gas, $\rho_\text{m}$; stars, $\rho_\star$;
galaxy bulge, $\rho_\text{b}$, and dark matter halo, $\rho_\text{dm}$.
Assuming that the diffuse gas has an exponential distribution in $Z$, 
$\rho_\dd= \rho_{0,\dd} \exp(-|Z|/h_\dd)$, equation~\eqref{eq:apg} can be integrated by 
parts,
\begin{equation}
  P\vert_{Z=0} =
 -h_\dd \rho_\dd(Z) \left.\deriv{\phi}{Z}\right\rvert^\infty_0 +
\int^\infty_0 h_\dd\rho_\dd(Z) \deriv{^2 \phi}{Z^2}\,\dd Z\,,
\end{equation}
and, since $\rho_\dd\vert_{Z\to\infty}=0$ and $\left.\partial{\phi}/\partial{Z}\right|_{Z=0}=0$, the mid-plane total gas pressure follows 
as
\begin{equation}
 P\vert_{Z=0} = 4\pi G \int^\infty_0 h_\dd\rho_\dd(Z)\sum_i\rho_i(Z)\, \dd Z \,.
  \label{eq:Prho}
\end{equation}
We assume that all the components associated with the galactic disc
(diffuse gas, molecular gas and stars) have exponential distributions in $Z$,
\begin{equation}
  \rho_j(\rcyl, Z) = \rho_{0,j}(\rcyl)\; e^{-|Z|/h_j(\rcyl)}
               = \frac{\Sigma_j(\rcyl)}{2\,h_j(\rcyl)} e^{-|Z|/h_j(\rcyl)}\,,
\end{equation}
where $h_j$ and $\Sigma_j$ are the scale height and surface density of 
the $j$-th component. Thus, for each of these disc components, we have
\begin{equation}
    \label{eq:P1disc}
P_{j}(\rcyl) = \frac{\pi}{2} G \Sigma_\dd(\rcyl) \Sigma_j(\rcyl)
  \frac{2h_\dd(\rcyl)}{h_j(\rcyl)+h_\dd(\rcyl)} \quad (\text{for }j=\dd,\text{m},\star)\,.
\end{equation}

To account for the contribution of the dark matter halo and stellar bulge,
we first examine the relation between their circular velocities -- discussed in the previous
section -- and the density profile. Since these two components are spherically symmetric,
the Poisson equation  can be written as (with $R$ the spherical radius)
\begin{equation}
 \frac{1}{\rsph^2}\deriv{}{\rsph}\left(\rsph^2\deriv{\phi_k}{\rsph}\right) = 4\pi G\rho_k
 \quad (\text{for }k=\text{dm,b})\,.
\end{equation}
We now recast the quantities in terms of the angular velocity
$\Omega = V/\rsph$, which is related to the gravitational potential as
$-\del\phi/\del\rsph = \Omega^2 r$,
\begin{equation}
 4\pi G\rho_k = \frac{1}{\rsph^2}\deriv{}{\rsph}\left(\rsph^3\Omega_k^2\right)
              = 3\Omega_k^2+2\Omega_k \rsph \deriv{\Omega_k}{\rsph}\,,
\end{equation}
then, for the dark matter halo and bulge components,
\begin{equation}
\rho_k(\rcyl,Z) = \frac{1}{2\pi G}
\left[\tfrac32  \Omega^2_k(\rcyl,Z) + \Omega_k(\rcyl,Z)\;S_k(\rcyl,Z)\right].
\end{equation}
Since we do not expect that the $\Omega$ and $S$ profiles (and thus, $\rho$) would vary
significantly for these two components over one diffuse gas scale height,
we approximate the integral \eqref{eq:Prho} to obtain the parts of the gas pressure 
that support the weights of the bulge and dark matter halo:
\begin{equation}
  P_{b/\text{dm}}(\rcyl) \approx 4\pi G h_d(\rcyl) \rho_{b/\text{dm}}(\rcyl,Z\!=\!0)
  \int^\infty_0\! \rho_d(Z') \dd Z' \;,\nonumber
\end{equation}
thus
\begin{align}
   \label{eq:P1bdm}
  P_{\text{b}/\text{dm}} \approx\; h_\dd(\rcyl) \Sigma_\dd(\rcyl) &\left[\tfrac32
  \Omega^2_{\text{b}/\text{dm}}(\rcyl,0)\right.\nonumber\\
  &\;\left.+\,\Omega_{\text{b}/\text{dm}}(\rcyl,0)\;S_{\text{b}/\text{dm}}(\rcyl,0)\right]\,.
\end{align}
Combining equations~\eqref{eq:P1disc} and \eqref{eq:P1bdm}, we obtain equation~\eqref{eq:P}.

\subsubsection{Gas pressure}
\label{ap:Pgas}
The gas pressure has several parts,
and the left-hand side of equation~\eqref{eq:P} has
\begin{equation}
  P = P_\text{t}+P_\text{th}+P_\text{m} +P_\text{cr}\,,
\end{equation}
where the thermal pressure is given by
\begin{equation}
  P_\text{th} = \gamma^{-1} \rho_\dd \cs^2\,,
\end{equation}
where $\gamma$ is the adiabatic index and $\cs$ is the sound speed; the turbulent pressure is
\begin{equation}
  \label{eq:Pturb}
  P_\text{t} = \tfrac{1}{3} \rho_\dd \vt^2\,,
\end{equation}
where $\vt$ is the turbulent velocity;
magnetic pressure due to both large-scale and random magnetic fields can be estimated as
\begin{equation}
  P_\text{m} = \xi P_\text{t}\,,
\end{equation}
with a certain constant $\xi$; and the cosmic ray pressure can be assumed to be 
proportional to the magnetic pressure,
\begin{equation}
  P_\text{cr} = \epsilon P_\text{mag}\,,
\end{equation}
with another constant $\epsilon$.

Assuming that the turbulent speed is equal to the sound speed,
the total pressure can be written as
\begin{equation}
 P(\rcyl) = \zeta \rho_\dd(\rcyl)v_0^2\, ,
\end{equation}
where 
\begin{equation}
 \zeta =\tfrac{1}{3}(1+\xi+\xi\epsilon)+\gamma^{-1}\,,
\end{equation}
which leads to equation~\eqref{eq:Pleft} with $\gamma = 5/3$ (monatomic ideal gas), 
$\mu=1$ (see Section~\ref{sec:ism}),  $\xi=1/4$,
$\epsilon=1$ (pressure balance), and then $\zeta = 1.1$.

\bsp
\label{lastpage}
\end{document}